%% file: ms.tex
\documentclass{article}

\usepackage[preprint, nonatbib]{neurips_data_2022}
\usepackage{pifont}% http://ctan.org/pkg/pifont
\newcommand{\xmark}{\ding{55}}%
\usepackage{xcolor}
\usepackage[utf8]{inputenc} % allow utf-8 input
\usepackage[T1]{fontenc}    % use 8-bit T1 fonts
\usepackage{hyperref}       % hyperlinks
\usepackage{url}            % simple URL typesetting
\usepackage{booktabs}       % professional-quality tables
\usepackage{amsfonts}       % blackboard math symbols
\usepackage{nicefrac}       % compact symbols for 1/2, etc.
\usepackage{microtype}      % microtypography
\usepackage{xcolor}         % colors
\usepackage{comment}
\usepackage{graphicx}
\usepackage{tabularx}
\usepackage{multirow}
\usepackage{amsmath}
\usepackage{wrapfig}
\usepackage{mdwlist}

\title{OLIVES Dataset: Ophthalmic Labels for Investigating Visual Eye Semantics}

\author{
  Mohit Prabhushankar$^1$, Kiran Kokilepersaud$^{1}$\thanks{Equal Contribution}  , Yash-yee Logan$^{1*}$,\\\textbf{Stephanie Trejo Corona$^{2*}$, Ghassan AlRegib$^1$, and Charles Wykoff$^2$}\\
  $^1$OLIVES at the Centre for Signal and Info. Processing, Georgia Tech, Atlanta, GA 30332, USA \\
  $^2$Retina Consultants Texas, Retina Consultants of America, Houston, Texas 77030, USA\\
  \texttt{\{mohit.p, kpk6, ylogan3, alregib\}@gatech.edu,}\\ \texttt{\{stephanie.trejo, ccwmd\}@retinaconsultantstexas.com}
}

\begin{document}

\onecolumn % make sure you keep this coverpage as one column. In this template, we force the coverpage to be one column with this command and then switch to double column for the remaining of the paper with the \doublecolumn command. 
\begin{basedescript}{\desclabelstyle{\pushlabel}\desclabelwidth{6em}}

\item[\textbf{Citation}]{M. Prabhushankar, K. Kokilepersaud, Y. Logan, S. Corona, G. AlRegib, and C. Wykoff "OLIVES Dataset: Ophthalmic Labels for Investigating Visual Eye Semantics" \emph{Advances in Neural Information Processing Systems} (NeurIPS 2022) Track on Datasets and Benchmarks, Nov 29 - Dec 1, 2022.}

\item[\textbf{Review}]{Data of Submission : 9 June 2022 \\ Date of Revision : 29 Aug 2022 \\ Date of Accept: 16 Sept 2022}

\item[\textbf{Dataset}]{\url{https://doi.org/10.5281/zenodo.7105232}}% If you do not have data related to this paper, you can remove the data keyword.

\item[\textbf{Dataset DOI}]{10.5281/zenodo.7105232}

\item[\textbf{Codes}]{\url{https://github.com/olivesgatech/OLIVES_Dataset}}% If you do not have data related to this paper, you can remove the data keyword.

\item[\textbf{Copyright}]{The authors grant NeurIPS a non-exclusive, perpetual, royalty-free, fully-paid, fully-assignable license to copy, distribute and publicly display all or parts of the paper. Personal use of this material is permitted. Permission from the authors must be obtained for all other uses, in any current or future media, including reprinting/republishing this material for advertising or promotional purposes, for resale or redistribution to servers or lists. }

\item[\textbf{Keywords}]{Ophthamology datasets, Biomarker analysis, Treatment prediction, Self-supervised learning}

\item[\textbf{Contact}]{\href{mailto:mohit.p@gatech.edu}{mohit.p@gatech.edu} OR \href{mailto:alregib@gatech.edu}{alregib@gatech.edu}\\ \url{https://ghassanalregib.info/} \\ }
\end{basedescript}

%Following command sequence was used to start the paper content from the following page and avoid numbering cover page.
\thispagestyle{empty}
\newpage
\clearpage
\setcounter{page}{1}

\maketitle

\begin{abstract}

\input{Sections/abstract}
\end{abstract}
\vspace{-1.5mm}
\section{Introduction}\label{sec:Intro}
\input{Sections/introduction}

\vspace{-1.5mm}
\section{Related Works}
\input{Sections/related_works}
\vspace{-1.5mm}
\section{OLIVES Dataset}
\input{Sections/dataset}

\vspace{-1.5mm}
\section{Clinical Applications}\label{sec:Results}
\input{Sections/results}
\vspace{-1.5mm}
\section{Discussion and Conclusion}\label{sec:Conclusion}
\vspace{-1.5mm}
\input{Sections/conclusion}
%%%%%%%%%%%%%%%%%%%%%%%%%%%%%%%%%%%%%%%%%%%%%%%%%%%%%%%%%%%%
\bibliographystyle{IEEEbib}
\bibliography{ref}
%\newpage

%%%%%%%%%%%%%%%%%%%%%%%%%%%%%%%%%%%%%%%%%%%%%%%%%%%%%%%%%%%%
%%%%%%%%%%%%%%%%%%%%%%%%%%%%%%%%%%%%%%%%%%%%%%%%%%%%%%%%%%%%
\section*{Checklist}

\begin{enumerate}

\item For all authors...
\begin{enumerate}
  \item Do the main claims made in the abstract and introduction accurately reflect the paper's contributions and scope?
    \answerYes{}
  \item Did you describe the limitations of your work?
    \answerYes{}. Please see Section~\ref{sec:Conclusion}
  \item Did you discuss any potential negative societal impacts of your work?
    \answerYes{}. Please see Section~\ref{sec:Conclusion}
  \item Have you read the ethics review guidelines and ensured that your paper conforms to them?
    \answerYes{}
\end{enumerate}

\item If you are including theoretical results...
\begin{enumerate}
  \item Did you state the full set of assumptions of all theoretical results?
    \answerNA{}
	\item Did you include complete proofs of all theoretical results?
    \answerNA{}
\end{enumerate}

\item If you ran experiments (e.g. for benchmarks)...
\begin{enumerate}
  \item Did you include the code, data, and instructions needed to reproduce the main experimental results (either in the supplemental material or as a URL)?
    \answerYes{}. Please see Section \ref{app: links}.
  \item Did you specify all the training details (e.g., data splits, hyperparameters, how they were chosen)?
    \answerYes{}. Please see training details for each experiment in Section \ref{subsec:Supervised}, \ref{subsec:Contrastive}, and \ref{subsec:Treatment_Bio}. Further details for the contrastive experiments can be found in Section \ref{app: sup_con}.
	\item Did you report error bars (e.g., with respect to the random seed after running experiments multiple times)?
    \answerYes{} See Tables found at \ref{tab:detection benchmarks} and \ref{tab:main_table}. Also, see Sections \ref{app: apps}, \ref{app: time-series}, and \ref{app: sup_con} for the standard deviation values associated with the time-series and contrastive learning experiments.
	\item Did you include the total amount of compute and the type of resources used (e.g., type of GPUs, internal cluster, or cloud provider)?
    \answerYes{}. Please see Section~\ref{app: cluster}
\end{enumerate}

\item If you are using existing assets (e.g., code, data, models) or curating/releasing new assets...
\begin{enumerate}
  \item If your work uses existing assets, did you cite the creators?
    \answerYes{}. Please see Section~\ref{app: attributions}.
  \item Did you mention the license of the assets?
    \answerYes{} Please see Section \ref{app: license}.
  \item Did you include any new assets either in the supplemental material or as a URL?
    \answerYes{} Please see Section \ref{app: links}. 
  \item Did you discuss whether and how consent was obtained from people whose data you're using/curating?
    \answerYes{}. Please see Section~\ref{subsec:Trials}, Line 138. 
  \item Did you discuss whether the data you are using/curating contains personally identifiable information or offensive content?
    \answerYes{}. Please see Section~\ref{subsec:Clinical}, Line 173.
\end{enumerate}

\item If you used crowdsourcing or conducted research with human subjects...
\begin{enumerate}
  \item Did you include the full text of instructions given to participants and screenshots, if applicable?
    \answerYes{} The clinical procedure for both trials is discussed in Section \ref{app: datasheet}.
  \item Did you describe any potential participant risks, with links to Institutional Review Board (IRB) approvals, if applicable?
    \answerYes{}. Please see Section~\ref{subsec:Biomarker}, Line 139 and Section~\ref{app: trail_description}, Line 1086.
  \item Did you include the estimated hourly wage paid to participants and the total amount spent on participant compensation?
    \answerNA{}
\end{enumerate}

\end{enumerate}

%%%%%%%%%%%%%%%%%%%%%%%%%%%%%%%%%%%%%%%%%%%%%%%%%%%%%%%%%%%%
\newpage

\appendix

%\end{document}
\newpage

%\appendix

\appendix

The appendix is divided into four sections. Appendix~\ref{app: Access} provides the dataset, labels, and benchmarking access. The dataset images and benchmarking codes are currently public, while the labels are provided privately as a link in Appendix~\ref{app: links}. Appendix~\ref{app: Stats} details some of the statistics of the dataset. This includes comparison against existing ophthalmology datasets, detailing the challenges within the OLIVES dataset, expanding on the full list of clinical labels that are available in the PRIME and TREX-DME clinical trials, and the exact procedure used to annotate the biomarkers. Appendix~\ref{app: Additional_labels} provides additional medical context to all the benchmarking results from Section~\ref{sec:Results}. Furthermore, experimental details including training setup, error bars, and computational resources are discussed. Finally, relevant procedural details regarding the PRIME and TREX DME clinical trials are discussed in Appendix~\ref{app: datasheet}, along with screenshots of relevant labels. 

\section{Dataset and Benchmarking Access}
\label{app: Access}
\input{Sections/appendixA}

\section{Dataset Statistics}\label{app: Stats}
\input{Sections/appendixB}

\section{Additional Results}\label{app: Additional_labels}
\input{Sections/appendixC}

\section{Datasheets}
\label{app: datasheet}
\input{Sections/appendixD}

\end{document}

%% file: Sections/abstract.tex
Clinical diagnosis of the eye is performed over multifarious data modalities including scalar clinical labels, vectorized biomarkers, two-dimensional fundus images, and three-dimensional Optical Coherence Tomography (OCT) scans. Clinical practitioners use all available data modalities for diagnosing and treating eye diseases like Diabetic Retinopathy (DR) or Diabetic Macular Edema (DME). Enabling usage of machine learning algorithms within the ophthalmic medical domain requires research into the relationships and interactions between all relevant data over a treatment period. Existing datasets are limited in that they neither provide data nor consider the explicit relationship modeling between the data modalities. In this paper, we introduce the Ophthalmic Labels for Investigating Visual Eye Semantics (OLIVES) dataset that addresses the above limitation. This is the first OCT and near-IR fundus dataset that includes clinical labels, biomarker labels, disease labels, and time-series patient treatment information from associated clinical trials. The dataset consists of $1268$ near-IR fundus images each with at least $49$ OCT scans, and $16$ biomarkers, along with $4$ clinical labels and a disease diagnosis of DR or DME. In total, there are $96$ eyes' data averaged over a period of at least two years with each eye treated for an average of $66$ weeks and $7$ injections. We benchmark the utility of \texttt{OLIVES} dataset for ophthalmic data as well as provide benchmarks and concrete research directions for core and emerging machine learning paradigms within medical image analysis.

%% file: Sections/introduction.tex
%%%%%%%%%%%%%%%%%%%%%%%%%%%%%%%%%%%%%%%%%%%%%%%%%
% Outline
%%%%%%%%%%%%%%%%%%%%%%%%%%%%%%%%%%%%%%%%%%%%%%%%

\begin{figure}[t]
\centering
\includegraphics[scale = .4]{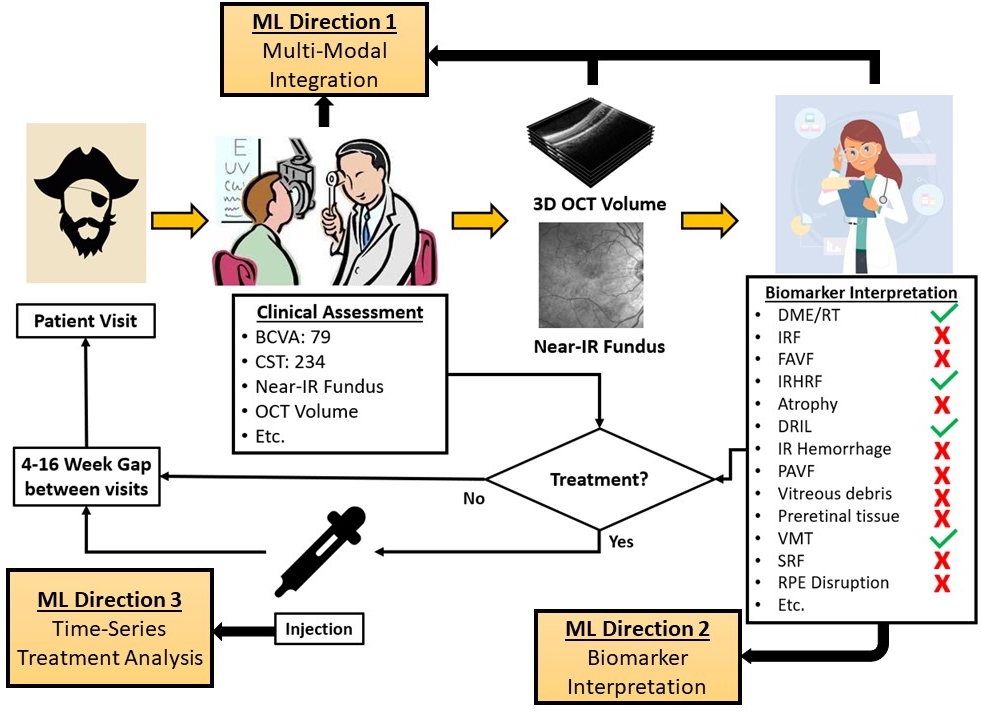}
\label{fig:clinical_process}\vspace{-1.5mm}
\caption{Complete summary of data collection process for the \texttt{OLIVES} dataset and potential research directions for the machine learning community.}\vspace{-1.5mm}
\end{figure}

% 1. Major Challenges of data Medicine
%  1 a)
\vspace{-1.5mm}
Ophthalmology refers to the branch of medical science that deals with the structure, functions, diseases, and treatments of the eye. A stylized version of the diagnostic and treatment process for a known disease is shown in Fig.~\ref{fig:clinical_process}. A patient's visit to a clinic is met with an assessment that includes visual acuity tests and collecting demographic information. This provides Best Corrected Visual Acuity (BCVA) scores, Patient ID, and Eye ID among other data. We term these as \emph{clinical labels}. Next, the patient undergoes diagnostic imaging that includes Fundus and OCT scans. Finally, a trained practitioner interprets the diagnostic scans for known \emph{biomarkers} for diseases. The authors in~\cite{golabbakhsh2013vessel} describe biomarkers as objective indicators of medically quantifiable characteristics of biological processes which are often diseases. The biomarkers along with the scans and clinical labels are used to assess the presence and severity of a patient's disease and a recommendation of a treatment is provided. If the recommendation is yes, the patient is treated and asked to visit again after a gap. 

A number of Machine Learning (ML) techniques have sought to either automate or interpret individual processes within Fig.~\ref{fig:clinical_process}. We annotate three such ML research directions within the pipeline for clinically aiding and monitoring disease diagnosis and treatment. The first direction involves assessing multi-modal data for clinical applications including predicting disease states. The second direction is an interpretation of biomarkers. Biomarkers act as intermediary data between medical scans and disease diagnosis that aid clinical reasoning. The last direction is analyzing time-series treatment data across the treatment period. This direction aids initial treatment prescription and patient monitoring. To the best of our knowledge, no existing dataset provides access to data that promotes all three stated research directions for the clinical process from Fig.~\ref{fig:clinical_process}. In this paper, we introduce the Ophthalmic Labels for Investigating Visual Eye Semantics (\texttt{OLIVES}) dataset that provides structured and curated data to promote holistic clinical research in ML for ophthalmic diagnosis.
\vspace{-1.5mm}
\paragraph{Clinical studies for \texttt{OLIVES} dataset} The \texttt{OLIVES} dataset is derived from the PRIME~\cite{hannah2021real} and TREX DME~\cite{payne2017randomized,payne2019randomized,payne2021long,wykoff2019intravitreal} clinical studies. Both the studies are prospective randomized clinical trials that were run between December 2013 and April 2021 at the Retina Consultants of Texas (Houston, TX, USA). Prospective trials refer to studies that evaluate the outcome of a particular disease during treatment. PRIME evaluates Diabetic Retinopathy (DR) and TREX-DME evaluates Diabetic Macular Edema (DME). The trials provide access to near-IR fundus images and OCT scans along with de-identified Electronic Medical Records (EMR) data of $96$ patients across an average of $66$ weeks. Biomarkers are retrospectively added to this data by experienced graders upon open adjudication.
\vspace{-1.5mm}
\begin{figure}[t!]
\centering
\includegraphics[scale = .4]{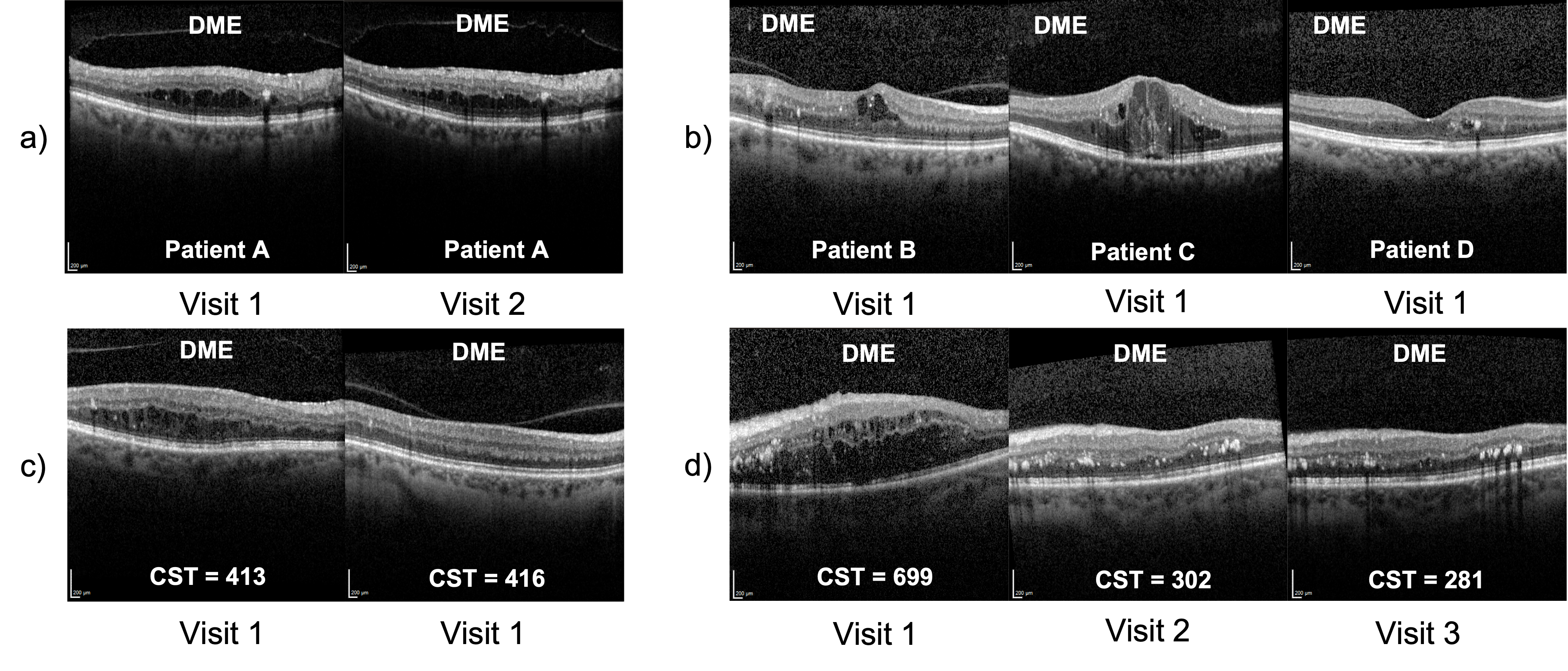}
\caption{An illustration of some challenges within the dataset. In a) variation in OCT belonging to the same patient at different visits is minimal. In b) varied disease manifestations are among different patients. c) At times CST values are very similar but the OCT is visually dissimilar. d) Shows that CST values gradually decrease as time progresses.}\vspace{-1.5mm}\label{fig:challenges}
\end{figure}
\paragraph{Challenging dataset for ML research} While challenges in natural images are generally contrived by intervening on top of data~\cite{temel2017cure, temel2018cure, chen2020action}, the complexities in ophthalmic datasets arise because of issues in data collection, inversion, representation and annotation.~\cite{cheng2020big}. \texttt{OLIVES} data modalities range from 1-dimensional numerical values (BCVA, Patient ID), vectorized biomarkers, 2-dimensional fundus images, and 3-dimensional scans (optical coherence tomography). Moreover, some of this data is objectively measured through instruments from patients (fundus, OCT), subjectively collected through eye tests (BCVA), while other data is interpreted and openly adjudicated through images (biomarkers). The variation within scans between visits can be minimal while the difference in manifestation of the same disease between patients may be substantial. This is shown in Fig.~\ref{fig:challenges}. The domain difference between OCT scans can arise due to pathology manifestation between patients (Fig.~\ref{fig:challenges}a and Fig.~\ref{fig:challenges}b), clinical labels (Fig.~\ref{fig:challenges}c), and the visit along the treatment process when the scan is taken (Fig.~\ref{fig:challenges}d). \texttt{OLIVES} provides access to these challenging data modalities that allow for innovative ML algorithms. 

\vspace{-1.5mm}
\paragraph{Contributions and significance of the dataset} The \texttt{OLIVES} dataset is curated to foster research in ophthalmic ML. The retrospective additions to the \texttt{OLIVES} dataset from its base clinical trials and its ensuing contributions include:
\begin{enumerate}
    \item Sixteen biomarker labels are added to the OCT scans of every first and last visit of all patients. We experimentally validate the necessity of biomarkers and provide benchmarks in Sections~\ref{subsec:Supervised} and~\ref{subsec:Treatment_Bio}. Along with biomarkers, \texttt{OLIVES} provides access to fundus, OCT scans, clinical labels and DR/DME diagnosis, thereby creating an ideal benchmarking mechanism for ophthalmic ML.
    \item We curate the clinical labels that have known correlations between the four data modalities. These include Best Corrected Visual Acuity (BCVA), Central Subfield Thickness (CST), Patient ID, and Eye ID. We demonstrate its utility for medically-grounded contrastive learning where augmentations are based on contrasting between clinical labels in Section~\ref{subsec:Contrastive}. Hence, \texttt{OLIVES} dataset promotes research in core and emerging ML paradigms. 
    \item The data and labels are made accessible to non-medical professionals. Biomarkers act as expert-annotated and interpretable visual indicators of diseases within OCT scans. The original labels from the clinical trials along with their data sheets are provided in Appendix~\ref{app: all_labels}. Additionally, an ML-specific set of labels which is relevant to the three mentioned research directions in Fig.~\ref{fig:clinical_process}, is provided in Appendix~\ref{app: ML_datasheet}.
\end{enumerate}

%% file: Sections/related_works.tex
\paragraph{Ophthalmology datasets} A number of publicly available ophthalmology datasets individually tackle each of the clinical modalities that exist in the \texttt{OLIVES} dataset. The authors in~\cite{khan2021global} provide a survey of $94$ existing open access ophthalmic datasets. Among $54$ of the $94$ datasets, the underlying data is that of fundus images. $19$ of the remaining datasets contain 3-dimensional OCT scans. The OCT scans provide structural information that enhances the performance of machine learning algorithms~\cite{khan2021global}. Only three of the $94$ considered open access datasets provide both OCT and fundus image modalities. The authors in~\cite{golabbakhsh2013vessel} provide $650$ OCT slices from a single volume. These are insufficient to leverage the data intensive machine learning algorithms to provide generalizable results. In contrast, the \texttt{OLIVES} dataset has $78,189$ slices taken from $1268$ volumes.~\cite{mahmudi2014comparison} provide OCT and fundus data from $50$ healthy patients. However, these are all for healthy eyes and disease manifestation is not observed. Other datasets including~\cite{kermany2018labeled} contains OCT scans for four OCT disease states: Healthy, Drusen, DME, and choroidal neovascularization (CNV).~\cite{farsiu2014quantitative} and~\cite{melinvsvcak2021annotated} introduced OCT datasets for age-related macular degeneration (AMD).~\cite{chiu2015kernel} contains OCT scans labeled with segmentation of regions with DME. However, these datasets do not possess comprehensive clinical information or a wide range of expert-annotated biomarkers. A complete overview that considers clinical labels, biomarkers, disease labeling, and time-series analysis is provided in Tables~\ref{tab:dataset_comparison} and~\ref{tab:dataset_comparison_time}. We refer to~\cite{khan2021global} to compare other statistics including number of image scans and applicability of existing datasets against \texttt{OLIVES}.  

\begin{comment}
\paragraph{Clinical labels and biomarkers in medical domain} The authors in~\cite{golabbakhsh2013vessel} describe biomarkers as objective indicators of medical state as observed and measured from outside the patient. They are quantifiable characteristics of biological processes. In this paper, the biological processes are diseases and biomarkers indicate the presence or absence of such diseases. Under limited circumstances, the authors in~\cite{golabbakhsh2013vessel} suggest that biomarkers can be surrogate endpoints in clinical trials. 
\end{comment}

\paragraph{Machine learning techniques on ophthalmic data} A number of works have separately addressed the research directions identified in Fig.~\ref{fig:clinical_process}. The authors in~\cite{temel2019relative} proposed transfer learning to screen for relative afferent pupillary defect due to lack of comprehensive data.~\cite{kermany2018identifying} showed that transfer learning methods could be utilized to classify OCT scans based on the presence of key biomarkers.~\cite{logan2022multimodal} introduced a dual-autoencoder framework with physician attributes to improve classification performance for OCT biomarkers.~\cite{de2018clinically} expanded previous work towards segmentation of a multitude of different biomarkers and referred for different treatment decisions. Other work has demonstrated the ability to detect clinical information from OCT scans which is significant for suggesting correlations between different domains. \cite{kawczynski2020development} showed that a model trained entirely on OCT scans could predict the associated BCVA value. Similarly \cite{arcadu2019deep} showed that values such as retinal thickness could be learned from retinal fundus photos. The \texttt{OLIVES} dataset provides a standardized benchmark to conduct research across applications, data modalities and machine learning paradigms.

% Please add the following required packages to your document preamble:
% \usepackage{booktabs}
% \usepackage{multirow}
\label{subsec:OLIVES data table}
\input{Tables/OLIVES}

%% file: Tables/OLIVES.tex
\begin{table}[h!]
\small
\caption{High-level overview of the \texttt{OLIVES} Dataset. The modality column details the type of data. The columns "Per Visit" and "Per Eye" indicate the amount of data in each modality on a respective visit or eye. $N_P$ is the number of visits that a patient $P$ takes to the clinic. The statistics across all eyes across all visits are shown in the Total Statistics column. Biomarkers are binary values, clinical labels are integers, fundus are 2D images, and OCT are 3D slices.}\label{tab:olives}
\begin{tabular}{@{}ccccc@{}}
\toprule
\multicolumn{5}{c}{OLIVES Dataset Summary}                        \\ \midrule
Modality & Per Visit & Per Eye & Total Statistics      & Overview \\ \midrule
\multirow{5}{*}{\begin{tabular}[c]{@{}c@{}}\\OCT\\ \\ Fundus\\ \\ Clinical\\ \\ Biomarker\end{tabular}} &
  \multirow{5}{*}{\begin{tabular}[c]{@{}c@{}}\\49\\ \\ 1\\ \\ 4\\ \\ 16\end{tabular}} &
  \multirow{5}{*}{\begin{tabular}[c]{@{}c@{}}\\$N_P$*49\\ \\ $N_P$\\ \\ $N_P$*4\\ \\ 1568\end{tabular}} &
  \multicolumn{1}{c|}{\multirow{5}{*}{\begin{tabular}[c]{@{}c@{}}\\78189\\ \\ 1268\\ \\ 5072\\ \\ 150528\end{tabular}}} &
  \multirow{5}{*}{\begin{tabular}[c]{@{}c@{}}\textbf{General:}\\ 96 Eyes, Visits every 4-16 weeks, \\ Average 16 visits and 7 injections/patient\\ \textbf{Clinical Labels obtained every visit:}\\ BCVA, CST, Patient ID, Eye ID\\ \textbf{Biomarkers labeled:}\\ IRHRF, FAVF, IRF, DRT/ME PAVF, VD, \\ Preretinal Tissue, EZ Disruption, IR Hemmorhages, \\ SRF, VMT, Atrophy, SHRM, RPE Disruption,\\ Serous PED\end{tabular}} \\
         &           &         & \multicolumn{1}{c|}{} &          \\
         &           &         & \multicolumn{1}{c|}{} &          \\
         &           &         & \multicolumn{1}{c|}{} &          \\
         &           &         & \multicolumn{1}{c|}{} &          \\
         &           &         & \multicolumn{1}{c|}{} &          \\
         &           &         & \multicolumn{1}{c|}{} &         \\
         &           &         & \multicolumn{1}{c|}{} &          \\
         &           &         & \multicolumn{1}{c|}{} &         \\
         &           &         & \multicolumn{1}{c|}{} &         
         \\\bottomrule
\end{tabular}
\end{table}

%% file: Sections/dataset.tex
\vspace{-1.5mm}
Statistics regarding the quantity of images and labels can be found in Table~\ref{tab:olives}. The \texttt{OLIVES} dataset is derived from the PRIME and TREX-DME trials. At every visit for each patient, ocular disease state data (DR/DME), clinical labels including BCVA, CST, Patient and Eye ID, and detailed ocular imaging including OCT, and fundus photography were obtained per the protocol in Section~\ref{app: all_labels}. This procurement of data continues across $N_P$ visits for every patient, where $N_P$ is the number of visits by a patient $P$. For instance, 3D longitudinal scans of the eye provide $49$ OCT scans per patient per visit. Across $N_P$ visits where $P$ can be any one of $96$ patients, the total number of OCT scans in the dataset is $78,189$. Note that on every visit, each patient undergoes testing to determine the requirement of a treatment per the clinical protocol described in Appendix~\ref{subsec:Trials}. Biomarkers are retrospectively added to each slice in the OCT scans for the first and last visits. Table~\ref{tab:olives} also indicates the total number of eyes, average number of visits and injections, and the time between visits.
\vspace{-1.5mm}
\begin{figure}[t!]
\centering
\includegraphics[scale = .4]{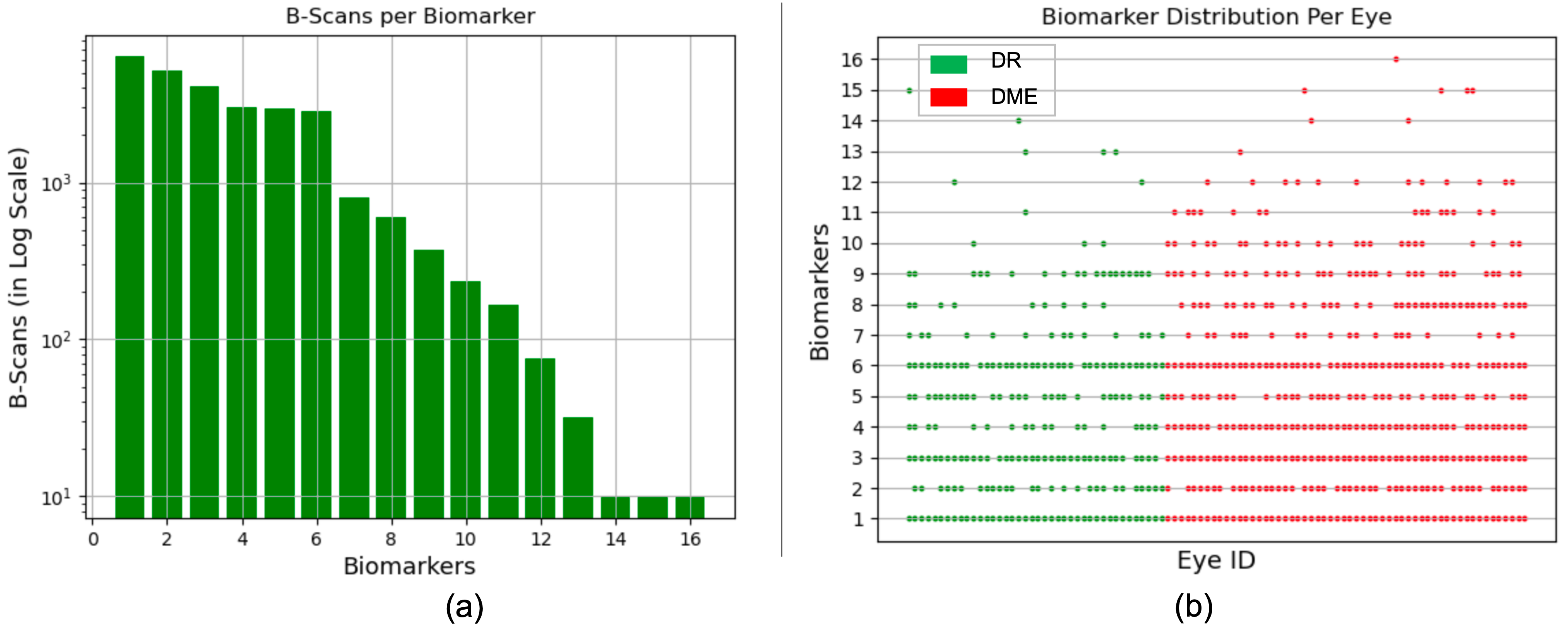}
\caption{(a) Histogram of the number of scans per biomarker. (b) Unique biomarkers per Eye ID.}\label{fig: eye_bio }\vspace{-1.5mm}
\end{figure}
\vspace{-1.5mm}
\subsection{Biomarker Generation}\label{subsec:Biomarker}
\vspace{-2mm}
After the clinical data collection process, we retrospectively provide additional insight into the OCT scans by providing corresponding biomarker labels. Biomarkers are quantifiable characteristics of biological processes in the eye. In this paper, the biological processes are diseases and biomarkers indicate the presence or absence of such diseases. Under limited circumstances, the authors in~\cite{golabbakhsh2013vessel} suggest that biomarkers can be surrogate endpoints in clinical trials. However, they caution against doing so unless the underlying clinical trial is specifically meant for the study. In both the PRIME and TREX DME studies, biomarkers are retrospectively labeled. As such, biomarkers may indicate the presence of diseases, but are not causal to these diseases. Hence, biomarkers are different from visual causal features from~\cite{prabhushankar2021extracting} or causal question-based analysis in~\cite{alregib2022explanatory} or causal factor analysis in~\cite{chalupka2014visual}. In the PRIME and TREX DME studies, images, clinical information, and biomarker labels were retrospectively collected at the Retina Consultants of Texas (Houston, TX, USA). This study was approved by the Institutional Review Board (IRB)/Ethics Committee and adheres to the tenets of the Declaration of Helsinki and Health Insurance Portability and Accountability Act (HIPAA). Informed consent was not required due to the retrospective nature of the study. A trained grader performed interpretation on OCT scans for the presence of $16$ different biomarkers including: intraretinal hyperreflective foci (IRHRF), partially attached vitreous face (PAVF), fully attached vitreous face (FAVF), intraretinal fluid (IRF), and diffuse retinal thickening or macular edema (DRT/ME). A full list of the biomarkers as well as their characteristics is provided in Section~\ref{app: biomarkers}. The full form of the abbreviations are given in Table~\ref{tab: abbreviation}. These biomarkers are chosen because of their visual attributes that correlate with presence or absence of disease states. The trained grader was blinded to clinical information whilst grading each of 49 horizontal OCT B-scans of both the first and last study visit for each individual eye. Open adjudication was done with an experienced retina specialist for difficult cases. In total, there are $9408$ OCT scans that consist of a $16\times 1$ biomarker vector where $1$ indicates the presence of the corresponding biomarker and a $0$ indicates its absence. We provide a histogram of the number of scans (\texttt{y-axis}) against their respective biomarkers in Fig.~\ref{fig: eye_bio }a. Note that the \texttt{y-axis} is in log-scale. We also depict the eye ID against the biomarkers in Fig.~\ref{fig: eye_bio }b. The green dots are eyes that indicate the presence of the corresponding biomarker on the \texttt{y-axis} that are diagnosed with DR. The red dots are for DME. It can be seen that a number of eyes have overlapping biomarkers even between diseases. Hence, biomarkers in isolation are insufficient to diagnose disease states, strengthening the case for multi-modal data.
\begin{figure}[t!]
\centering
\includegraphics[scale = .4]{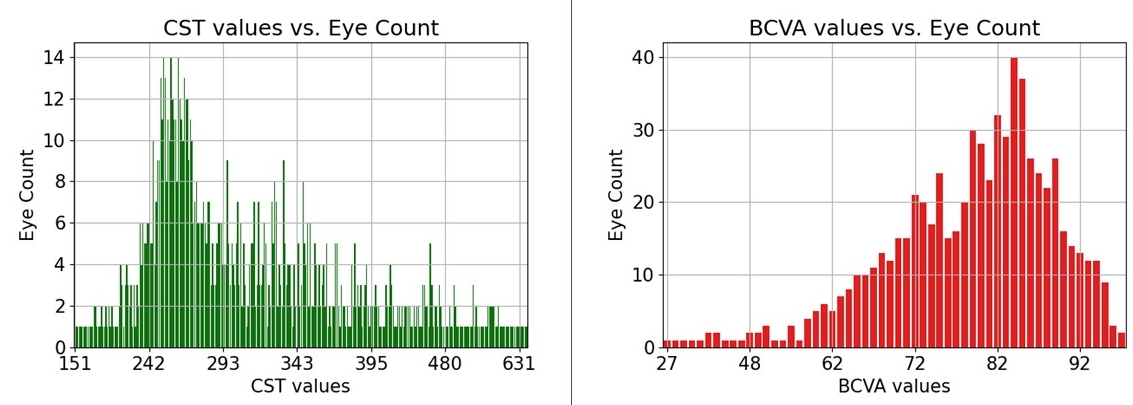}
\caption{Distribution of CST and BCVA labels in \texttt{OLIVES} dataset based on number of eyes  associated with each clinical value.}\label{fig: distribution}
\end{figure}
\vspace{-1.5mm}
\subsection{Clinical Labels}\label{subsec:Clinical}
\vspace{-1.5mm}
Within the \texttt{OLIVES} dataset, we have explicit clinical information regarding the Best Central Visual Acuity (BCVA), Central Subfield Thickness (CST), and identity of the eye. ETDRS best-corrected visual acuity (BCVA) is a visual function assessment performed by certified examiners where a standard vision chart is placed 4-meters away from the patient. The patient is instructed to read the chart from left to right from top to bottom until the subject completes 6 rows of letters or the subject is unable to read any more letters. The examiner marks how many letters were correctly identified by the patient. Central subfield thickness (CST) is the average macular thickness in the central 1-mm radius of the ETDRS grid. Both BCVA and CST are coarse measurements over the eye as opposed to Biomarkers that exist for fine-grained longitudinal slices of the eye. BCVA can range from $0-100$ and CST from $100-1300$. We show in Fig.~\ref{fig: distribution} the number of eyes (\texttt{y-axis}) that have the associated value (\texttt{x-axis}) for both BCVA and CST. This graph shows that our dataset has a wide variation in terms of range of clinical values across a multitude of eyes in the dataset. This is advantageous as it shows the dataset is not biased to any specific range of values or localized to single eye instances. A full list of all clinical labels present in PRIME and TREX-DME clinical trials are shown in Section~\ref{app: all_labels}. No personally identifiable information was included in compliance with HIPAA regulations.
\begin{figure}[h!]
\centering
\includegraphics[scale = .4]{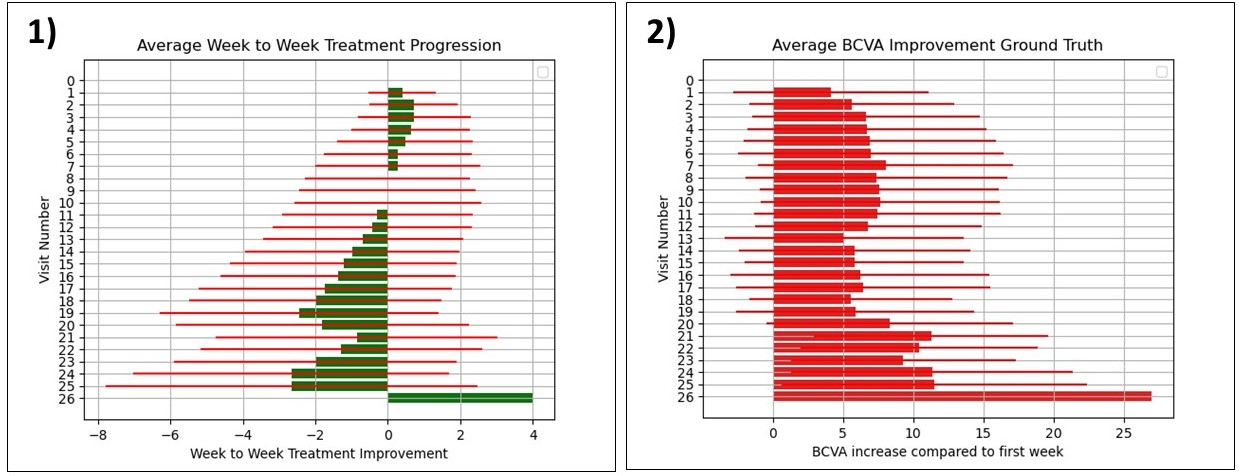}
\caption{1) A plot of average number of visits by patients that were an improvement or deterioration from previous week. Red bars indicate the standard deviation across all patients. 2) Plot of average change in BCVA with respect to the first week.}\label{fig: progression}
\end{figure}
\vspace{-1.5mm}
\subsection{Time-series data}
\vspace{-1.5mm}
A core novelty of the dataset is that data exists for each patient visit across a defined period of time. As a result, it is possible to analyze trends in the collected imaging and clinical data over the visiting period of the patient. This is shown in Fig.~\ref{fig: progression} with an overall progression analysis shown by the bar graphs. Graph 1 indicates whether on average there was an improvement from the previous visit. This was computed by assigning a value of 1 for improvements and -1 for deterioration. This is accumulated every visit and the average across all patients is calculated on a per visit basis. From this plot, it can be observed that in the dataset, eyes generally improve on every visit until about the tenth visit. However, graph 2 n Fig.~\ref{fig: progression} shows that while visit to visit improvement declines, the overall improvement when compared with the first visit is generally substantial. This graph was computed by taking the difference between the current visit's BCVA and first visit's BCVA and averaging across all patients. The statistics of the number of patients every visit and visualizations of patient treatment is shown in Figs.~\ref{fig: patient_dist} and~\ref{fig: progression_app} respectively.

\subsection{Interaction between Data Modalities}
\label{subsec: Interaction_data}
Clinical labels correspond to measurements that pertain to the entire visual system, including the visual mechanism in the eye. These measurements give an overview of the health of the eye, but they do not enable fine-grained analysis of structures that exist within the eye. Biomarker labels exist at the longitudinal slice level. They are detailed labels for every slice of the eye and provide a fine-grained analysis of the biological structures that exist within the eye. Clinical studies such as~\cite{hannouche2012correlation} and~\cite{sun2014disorganization} suggest that measured clinical labels can act as indicators of structural changes that manifest themselves in OCT scans and fundus images as well as the severity of disease associated with the patient. For example, visually, it can be observed that OCT scans with the same BCVA values exhibit more common structural characteristics than scans with different BCVA values. Furthermore, all data modalities exhibit visual, structural and clinical changes across the treatment period. \texttt{OLIVES} dataset allows for exploiting these correlations between OCT, fundus, clinical labels, biomarkers, diseases, and treatment states.

%% file: Sections/results.tex
\vspace{-3mm}
With the multitude of modalities that exist within the \texttt{OLIVES} dataset, there is potential for research in a wide variety of ML applications. Within this section, we focus on applications, and benchmarks, that showcase key features of the dataset identified from Fig.~\ref{fig:clinical_process}, but acknowledge that other novel setups and formulations of the problem are possible and intended. These applications include multi-modal integration of OCT scans and biomarker/clinical labels, biomarker detection and interpretation using contrastive learning, and time-series treatment analysis. 

\subsection{Multi-Modal Integration Between OCT and Biomarkers/Clinical Labels}
\label{subsec:Supervised}

\input{Tables/DRDME}

\paragraph{Baseline Detection of DR/DME with OCT} 
Since biomarkers are only available for the first and last clinical visits, we use the corresponding OCT at those visits for this baseline analysis. The entire dataset is partitioned by eyes into train, test and validation splits. Additional details about train/test/validation splits is in Appendix \ref{app: multi-modal details}.  We evaluate performance with balanced accuracy, precision and recall performance metrics. The results for the baseline OCT model is shown in the first row of Table \ref{tab:detection benchmarks}. Additional results showing specificity and sensitivity are in Table \ref{tab:detection benchmarks appendix} in the Appendix. This and subsequent experiments are conducted using multiple random seeds for DR/DME detection and an average score and standard deviation is reported for balanced accuracy.
\vspace{-1.5mm}
\paragraph{Supervised Learning with Clinical Labels}
We aim to use clinical labels as an additional modality to aid the baseline model. However, to determine the suitability of this auxiliary data type, we first evaluate its impact on the classification of DR and DME. To do this we first find all unique clinical labels present in the dataset with their associated disease labels. Then, we create a training set with $70\%$ of these clinical labels along with test and validation sets of $20\%$ and $10\%$ proportions respectively. This yields $1107$ unique clinical labels for training, $306$ for testing and $122$ for validation. Within the test set, half the samples are DR and the remaining DME. The second row on Table \ref{tab:detection benchmarks} shows that CST and BCVA used as clinical features are more effective than the unimodal OCT baseline for DR/DME detection.

% Similarly, there is a need to evaluate the discriminative power of clinical labels at differentiating between classes. Therefore, we perform a similar analysis as previously described using BCVA and CST clinical labels as features to characterize the diseases. There are $1,535$ unique clinical label features among $1107$, $306$, $122$ samples which are used for train, test and validation sets respectively. Half the samples within the test set belong to DR and the remaining to DME. We train an ANN with two linear layers and Relu activation between. The second row on Table \ref{tab:detection benchmarks} shows that clinical labels are not as effective as biomarkers at distinguishing between DR and DME. However, CST and BCVA used as clinical features are more effective than the unimodal OCT baseline.   
\vspace{-1.5mm}
\paragraph{Supervised Learning with Biomarkers}
We perform a similar analysis as described in supervised learning with clinical labels but using biomarkers as features. Hence, we substitute the clinical labels with biomarkers to characterize the diseases. There are $286$ unique biomarker label features among which $200$, $58$, $28$ samples are used for train, test and validation sets respectively. From the third row in Table \ref{tab:detection benchmarks}, we observe that using biomarkers on their own leads to a $9.72\%$ increase in DR and DME classification over baseline results. 

% We aim to use biomarkers as an additional modality to aid the baseline model. However, to determine the suitability of this auxiliary data type, we first evaluate its impact on the classification of DR and DME. To do this we first find all unique biomarkers present in the dataset with their associated disease labels. Then, we create a training set with $70\%$ of these biomarkers along with test and validation sets of $20\%$ and $10\%$ proportions respectively. This yields $200$ unique biomarkers for training, $58$ for testing and $28$ for validation. These are used to train a shallow artificial neural network (ANN) with four linear layers and LeakyRelu activation between. Biomarker features are normalized to zero mean and unit standard deviation. Within the test set,  $23$ samples are DR and the remaining DME. From the third row in Table \ref{tab:detection benchmarks}, we observe that using biomarkers on their own leads to a $9.72\%$ increase in DR and DME classification. 
\vspace{-1.5mm}
\paragraph{Multi-Modal Learning with OCT and Clinical Labels}
Having seen that clinical labels are more effective than the baseline model at DR/DME classification, we now investigate how to use the clinical label modality to aid the OCT model. Clinical labels and OCT are independently given as input to their models as described previously. We optimize both models jointly with a loss function that allows knowledge, in the form of logits, from the clinical model to guide the optimization of the OCT model. A detailed description of this optimization scheme can be seen in Appendix \ref{app: guided loss}. During testing, only the OCT model, having been optimized jointly with the other model, is used to classify the disease states. The fourth row of Table~\ref{tab:detection benchmarks} shows that clinical labels also aid the OCT model at characterizing the diseases albeit not the most effectively. 

% In like manner, we investigate the impact that clinical labels as features can have on aiding an OCT model to classify the two diseases. Again, the same train, test and validation splits are used in this analysis and each model is fed their input modality and optimized jointly using the same loss detailed in Appendix \ref{app: guided loss}. During testing, only the OCT model is used for inference. The forth row of Table~\ref{tab:detection benchmarks} shows that clinical labels also aid the OCT model at characterizing the diseases albeit not as effectively as the OCT and biomarker model. 
\vspace{-1.5mm}
\paragraph{Multi-Modal Learning with OCT and Biomarkers}
In like manner, we investigate the impact that biomarkers as features can have on aiding an OCT model to classify the two diseases. Each model is fed their input modality and optimized jointly using the same loss detailed in Appendix \ref{app: guided loss}. During testing, only the OCT model is used for inference. The final row in Table \ref{tab:detection benchmarks} shows that this is the most effective technique that significantly improves all baseline classification metrics. 
\vspace{-1.5mm}

% Having seen that biomarkers are more effective than the baseline model at DR/DME classification, we now investigate how to use the biomarker modality to aid the OCT model. We use the same train, test and validation split as the baseline OCT model and the biomarkers associated with each B-scan. Biomarkers and OCT are independently given as input to their models described previously. We optimize both models jointly with a loss function that allows knowledge, in the form of logits, from the biomarker model to guide the optimization of the OCT model. A detailed description of this optimization scheme can be seen in Appendix \ref{app: guided loss}. During testing, only the OCT model, having been optimized jointly with the other model, is used to classify the disease states. The final row in Table \ref{tab:detection benchmarks} shows that this is the most effective technique that significantly improves all baseline classification metrics. 
\vspace{-1.5mm}

\subsection{Biomarker Interpretation with Contrastive Learning}
\label{subsec:Contrastive}
\vspace{-1.5mm}
\input{Tables/contrastive_learning}
Due to the prohibitive costs of expert-annotated biomarker labels, contrastive learning~\cite{le2020contrastive, chen2020simple, prabhushankar2021contrastive} approaches have garnered attention because of their state of the art self-supervised performance. These approaches generally create a representation space through minimizing the distance between positive pairs of images and maximizing the distance between negative pairs. Traditional approaches, like SimCLR ~\cite{chen2020simple}, generate positives from augmentations of a single image and treat all other images in a batch as negatives. More modern approaches like Moco v2 ~\cite{chen2020improved} incorporate a queue system for additional negative samples while extensions of this include PCL ~\cite{li2020prototypical} that introduce a clustering approach on the representation space. While these approaches have shown promising results on natural images, such augmentations are unrealistic for medical images that rely on fine-grained changes within OCT scans to detect diseases. Instead, we propose using clinically relevant labels as a means to better choose positive pairs. Since \texttt{OLIVES} provides a larger pool of clinical labels than biomarker labels, this task fits well within the scope of the dataset. Hence, \texttt{OLIVES} enables research into novel and multi-modal contrastive learning strategies. We implement one such strategy through reformulating the supervised contrastive loss \cite{khosla2020supervised} in a clinical context as discussed in \cite{kokilepersaud2022supcon} with related work located at \cite{kokilepersaud2022gradient, kokilepersaud2022volumetric}. Implementation details are provided in Section~\ref{app: sup_con}. 
\begin{figure}[t]
\small
\centering
\includegraphics[scale = .35]{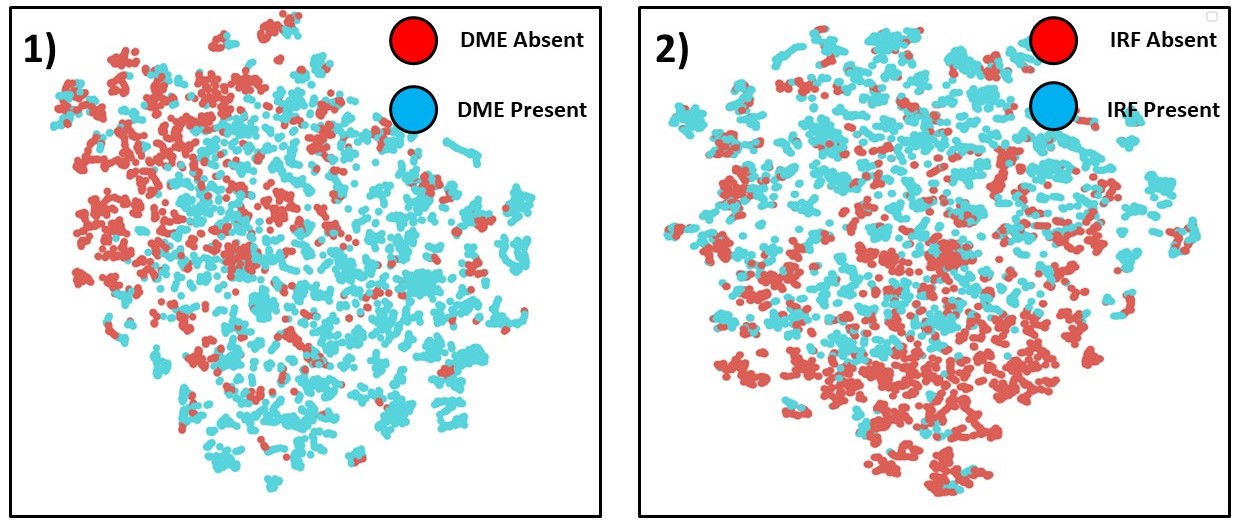}
\caption{T-SNE visualization of the OLIVES Biomarker test set labeled by the presence or absence of DME and IRF. We can effectively achieve an embedding space that is separable with respect to biomarkers 1) DME and 2) IRF.} 
\label{fig: dme_embed}
\end{figure} 

We train a Resnet-18~\cite{he2016deep} encoder with the clinically labeled data using the clinically aware supervised contrastive loss. After training the encoder with supervised contrastive loss, we freeze the weights of the encoder and append a linear layer to its output. This linear layer is trained using cross-entropy loss to distinguish between the presence or absence of the biomarker of interest in the OCT scan. Not all of these biomarkers exist in sufficiently balanced quantities to train a model to identify their presence or absence within an image. Hence, we use five biomarkers that fit this criteria in this study. The training details are presented in Section~\ref{app: sup_con}. We compare this method with three state of the art self-supervised algorithms in Table~\ref{tab:main_table}. We evaluate performance in terms of individual biomarker accuracy and f1-score as well as in the setting where the goal is to simultaneously perform a multi-label classification of biomarkers. Performance is measured by average AUROC, specificity, and sensitivity. We observe that a training strategy that chooses positives based on the clinical data Eye ID, BCVA, and CST outperforms baseline self-supervised methods in both a multi-label classification task as well as individual biomarker detection performance. While the results in Section~\ref{subsec:Supervised} make use of correlations between the biomarkers and clinical labels with disease states, these results depict the correlations between the label modalities.

In Figure \ref{fig: dme_embed}, we visualize the test set t-SNE embeddings of two different biomarkers from a model trained using the BCVA clinical label. We observe that even without any fine-tuning on the actual biomarker label of interest, we are able to get an embedding space where the absence and presence of DME and IRF form distinct clusters. This gives credence to the idea that there exists relationships between the biomarker and clinical label domains as training on only clinical labels leads to a separable space within the biomarker domain.
\vspace{-1mm}
\subsection{Time-Series Treatment Analysis}
\label{subsec:Treatment_Bio}
\vspace{-1mm}
% Additionally, we present two experimental manifestations based off the temporal nature of the data.
The multi-modal nature of \texttt{OLIVES} dataset allows for a large combination of experimental setups to analyze treatments. We present two experimental manifestations based off the temporal nature of the data: a) Predicting visit-by-visit successive treatment effects and b) Predicting the final ocular state using Biomarkers. A key metric used to evaluate treatment progression or regression is BCVA. At each visit to the clinic, patients' ocular disease states are evaluated and BCVA and other clinical labels are recorded. From a machine learning perspective, this motivates an analysis of treatment effect over consecutive weeks to predict how BCVA scores will change based on the state of the eye captured via OCT or Fundus. We detail the exact experimental procedure in Appendix~\ref{app: time-series}. We evaluate the performance of this strategy on both fundus images and 3D OCT volumes. We use a Resnet-18 \cite{he2016deep}, ResNet-50 \cite{he2016deep}, DenseNet-121 \cite{huang2017densely}, EfficientNet \cite{tan2019efficientnet}, and Vision Transformer~\cite{dosovitskiy2010image} (using a patch size of 32, 16 transformer blocks, 16 heads in multi-attention layer). For the OCT volumes, we utilize a version of each architecture that uses three-dimensional convolution layers. Performance in both modalities is reported in Table \ref{tab:bcva_pred}. We observe that the model is able to learn distinguishing features between the two classes, with better performance when using the OCT volumetric data. Additionally, we present results for predicting the final state of $16\times 1$ biomarker vector given the initial biomarker vector for individual patients in Fig.~\ref{fig:final_bio_pred}. Similar to the week-wise case, these results indicate correlation among multiple modalities as well as the ability of ML algorithms to predict ocular states given treatment.

\input{Tables/treatment_effect}

%% file: Tables/DRDME.tex
\begin{table}[h!]  
\small
  \begin{center}
    \caption{Benchmark results for DR/DME detection.}
    \label{tab:detection benchmarks}
    \begin{tabular}{*{16}{c c c c c}}
      \toprule % <-- Toprule here
      \multirow{2}{*}{\textbf{Experiments}}  &
      \multirow{2}{*}{\textbf{Model}}  &
      \multirow{2}{*}{\textbf{Balanced Accuracy}} &
    %   \multicolumn{2}{c}{\textbf{Precision}} &
      \multirow{2}{*}{\textbf{Specificity}}
        & \multirow{2}{*}{\textbf{Sensitivity}} \\
      & & & & & & \\
      \hline
      \midrule % <-- Midrule here
      OCT & R-18 & 70.15\% $\pm$ 4.69 & 0.608 & 0.794 \\
      Clinical & MLP & 75.49\% $\pm$ 1.98 & 0.758 & 0.751 \\
      Biomarker & MLP & 79.87\% $\pm$ 3.03 & 0.826 & 0.771 \\
      OCT + Clinical & R-18 + MLP & 75.92\% $\pm$ 3.05 & 0.566 & 0.952 \\
      OCT + Biomarker & R-18 + MLP & 82.33\% $\pm$ 3.59 & 0.742 & 0.904 \\
      \bottomrule % <-- Bottomrule here
    \end{tabular}
  \end{center}
\end{table}

%% file: Tables/contrastive_learning.tex
% Please add the following required packages to your document preamble:
% \usepackage{booktabs}
% \usepackage{multirow}
% \usepackage{graphicx}
\begin{table}[t!]
\centering
\caption{Benchmark of the performance of supervised contrastive training on images with clinical and biomarker data. The standard deviations are shown in Table~\ref{tab:main_std}.}
\label{tab:main_table}
\resizebox{\textwidth}{!}{%
\begin{tabular}{@{}cccccccccccccc@{}}
\toprule
\multirow{3}{*}{Method} &
  \multicolumn{10}{c}{Biomarkers} &
  \multicolumn{3}{c}{Metrics} \\ \cmidrule(l){2-14} 
 &
  \multicolumn{2}{c}{IRF} &
  \multicolumn{2}{c}{DRT/ME} &
  \multicolumn{2}{c}{IRHRF} &
  \multicolumn{2}{c}{FAVF} &
  \multicolumn{2}{c}{PAVF} &
  \multicolumn{1}{c}{\multirow{2}{*}{AUROC}} &
  \multicolumn{1}{c}{\multirow{2}{*}{\begin{tabular}[c]{@{}c@{}}Average \\ Specificity\end{tabular}}} &
  \multirow{2}{*}{\begin{tabular}[c]{@{}c@{}}Average \\ Sensitivity\end{tabular}} \\
 &
  \multicolumn{1}{c}{Accuracy} &
  \multicolumn{1}{c}{F1-Score} &
  \multicolumn{1}{c}{Accuracy} &
  \multicolumn{1}{c}{F1-Score} &
  \multicolumn{1}{c}{Accuracy} &
  \multicolumn{1}{c}{F1-Score} &
  \multicolumn{1}{c}{Accuracy} &
  \multicolumn{1}{c}{F1-Score} &
  \multicolumn{1}{c}{Accuracy} &
  \multicolumn{1}{c}{F1-Score} &
  \multicolumn{1}{c}{} &
  \multicolumn{1}{c}{} &
   \\ \hline \midrule
\multicolumn{1}{c}{PCL \cite{li2020prototypical}} &
  \textbf{76.50\%} &
  \multicolumn{1}{c}{0.717} &
  80.11\% &
  \multicolumn{1}{c}{0.761} &
  59.10\% &
  \multicolumn{1}{c}{0.683} &
  76.30\% &
  \multicolumn{1}{c}{0.773} &
  51.40\% &
  \multicolumn{1}{c}{0.165} &
  \multicolumn{1}{c}{0.767} &
  \multicolumn{1}{c}{0.741} &
  0.604 \\
\multicolumn{1}{c}{SimCLR \cite{chen2020simple}} &
  75.13\% &
  \multicolumn{1}{c}{0.716} &
  80.61\% &
  \multicolumn{1}{c}{0.772} &
  59.03\% &
  \multicolumn{1}{c}{0.675} &
  75.43\% &
  \multicolumn{1}{c}{0.761} &
  52.69\% &
  \multicolumn{1}{c}{0.249} &
  \multicolumn{1}{c}{0.754} &
  \multicolumn{1}{c}{0.747} &
  0.614 \\
\multicolumn{1}{c}{Moco V2 \cite{chen2020improved}} &
  76.00\% &
  \multicolumn{1}{c}{\textbf{0.720}} &
  82.24\% &
  \multicolumn{1}{c}{0.793} &
  59.60\% &
  \multicolumn{1}{c}{0.692} &
  75.00\% &
  \multicolumn{1}{c}{0.784} &
  52.69\% &
  \multicolumn{1}{c}{0.211} &
  \multicolumn{1}{c}{0.770} &
  \multicolumn{1}{c}{0.762} &
  0.651 \\ \hline \midrule
\multicolumn{1}{c}{Eye ID} &
  72.63\% &
  \multicolumn{1}{c}{0.674} &
  80.20\% &
  \multicolumn{1}{c}{0.778} &
  58.00\% &
  \multicolumn{1}{c}{0.674} &
  74.93\% &
  \multicolumn{1}{c}{0.725} &
  \textbf{65.56\%} &
  \multicolumn{1}{c}{\textbf{0.588}} &
  \multicolumn{1}{c}{0.767} &
  \multicolumn{1}{c}{\textbf{0.776}} &
  0.656 \\
\multicolumn{1}{c}{CST} &
  75.53\% &
  \multicolumn{1}{c}{0.720} &
  \textbf{83.06\%} &
  \multicolumn{1}{c}{\textbf{0.811}} &
  \textbf{64.30\%} &
  \multicolumn{1}{c}{\textbf{0.703}} &
  76.13\% &
  \multicolumn{1}{c}{0.766} &
  62.16\% &
  \multicolumn{1}{c}{0.509} &
  \multicolumn{1}{c}{\textbf{0.790}} &
  \multicolumn{1}{c}{0.772} &
  \textbf{0.675} \\
\multicolumn{1}{c}{BCVA} &
  74.03\% &
  \multicolumn{1}{c}{0.701} &
  80.27\% &
  \multicolumn{1}{c}{0.770} &
  58.8\% &
  \multicolumn{1}{c}{0.672} &
  \textbf{77.63\%} &
  \multicolumn{1}{c}{\textbf{0.785}} &
  58.06\% &
  \multicolumn{1}{c}{0.418} &
  \multicolumn{1}{c}{0.776} &
  \multicolumn{1}{c}{0.713} &
  0.645 \\ \bottomrule
\end{tabular}%
}
\end{table}

%% file: Tables/treatment_effect.tex
% Please add the following required packages to your document preamble:
% \usepackage{booktabs}
% \usepackage{graphicx}
\begin{table}[h!]
\centering
\caption{Benchmark Performance of predicting treatment effects from time-series Fundus and OCT data.}
\label{tab:bcva_pred}
\resizebox{0.7\textwidth}{!}{%
\begin{tabular}{@{}ccccc@{}}
\toprule
Model & Image Modality & Accuracy & Precision & Recall \\ \midrule
\multirow{2}{*}{\centering ResNet-18} & Fundus         & 55.19\% $\pm$ 10.9  & 0.256     & 0.343  \\
 & OCT Volume     & 57.59\% $\pm$ 9.51  & 0.359     & 0.326  \\
\cline{2-5}
\multirow{2}{*}{\centering ResNet-50} & Fundus         & 48.73\% $\pm$ 13.3  & 0.372     & 0.3296  \\
& OCT Volume     & 57.70\% $\pm$ 9.1  & 0.301     & 0.1826  \\
\cline{2-5}
\multirow{2}{*}{\centering DenseNet-121} & Fundus         & 53.00\% $\pm$ 8.9  & 0.273     & 0.259  \\
& OCT Volume     & 54.75\% $\pm$ 4.92  & 0.219     & 0.188  \\
\cline{2-5}

\multirow{2}{*}{\centering EfficientNet} & Fundus         & 56.06\% $\pm$ 4.85  & 0.292     & 0.217  \\
 & OCT Volume     & 60.65\% $\pm$ 4.09  & 0.3613     & 0.1633  \\
 \cline{2-5}
ViT & Fundus & 55.01\% $\pm$ 3.27 & 0.285 & 0.350 \\
\bottomrule
\end{tabular}%
}
\end{table}

%% file: Sections/conclusion.tex
\paragraph{Domain Difference and Adaptation in Multi-Modal Data}
The data in \texttt{OLIVES} is derived from two studies. As mentioned in Section~\ref{sec:Intro}, the domain difference in ophthalmic data can arise from sources such as treatment, disease manifestation, and clinical labels. In natural images, one source of domain difference is the equipment used for imaging. In PRIME and TREX studies, the same imaging and grading modalities, the Heidelberg Spectralis HRA+OCT software, is used in the same clinic. We provide extensive experiments in Appendix~\ref{app: domain-Shift} and~\ref{app: domain adaptation} to characterize possible domain differences on OLIVES. In Table~\ref{tab:Trial_domain}, we show that the biomarker detection results when trained and tested on PRIME trial is lower than when trained with TREX and tested on PRIME. This is because a longer treatment period on TREX dataset provides more diverse data that is conducive for training ML algorithms. Intuitively, this suggests that treatment causes domain shift in data, which is illustrated in Table~\ref{tab:Treatment_domain}. Training and testing within the first week data provides the best results for biomarker detection. This analysis is further expanded in Fig.~\ref{fig:domainadapt}. Rather than showing domain difference, we adapt between the first and last visit domains. Specifically, we use a part of the last visit data to train with the first visit data and show that: a) adapting between OCT scans before and after treatment is possible, and b) the addition of biomarkers increases the results for diagnosis of DR/DME. Hence, \texttt{OLIVES} provides data modalities that promotes research in treatment-based domain difference and adaptation in medical data.

\paragraph{Dataset Limitations, Societal Impact, and Ethical Concerns} The \texttt{OLIVES} dataset is derived from two clinical studies conducted from only one U.S. clinic. While there is a range in the age, ethnicity and racial demographics within the cohorts, this range is only limited to one geographical location. Hence, an end-to-end system can be biased. To mitigate this limitation, we provide links to existing open access ophthalmic datasets in Appendix~\ref{app: limitations} that are collected from other parts of the world. While none of these datasets are as rich as our own in terms of numbers, modalities, or labels, they can be used to modularly test algorithms. We present one such result in Table~\ref{tab:severe_table} and show that combining datasets allows for higher results. The PRIME and TREX trials are randomized clinical studies with the goal of comparing different treatment regimens. These studies aim to find the best practices for how and when they should treat patients to get the most optimal outcomes. However, there are no control groups within the studies that did not receive treatment. While this is common in clinical trials~\cite{wykoff2019intravitreal}, it adds a new challenge to ML-focused research of time-series analysis. We list datasets that provide healthy images in Appendix~\ref{app: limitations} to complement \texttt{OLIVES}. We believe that a combination of datasets taken over multiple geographical regions, times, and disease states is essential to construct generalizable and ethical ML models. ML models can potentially amplify existing inequalities within healthcare access~\cite{chen2021ethical}. For instance, the data in \texttt{OLIVES} is collected from December 2013 to April 2021, which implies the participants had the time and means to be part of these trials. This may not always be the case for disadvantaged groups. Hence, any benefit that machine learning could provide will be restricted to small subsets of society unless thought is put into preventing this disparity. Hence, a careful analysis of potential concerns is required to use \texttt{OLIVES} and any other dataset to enrich the functionality and adaptability of machine learning algorithms in everyday lives.

\paragraph{Conclusion} We introduce the \texttt{OLIVES} dataset to bridge the gap between existing ophthalmic datasets and the clinical diagnosis and treatment process. \texttt{OLIVES} provides curated and contained data that can be used for clinical interpretation of biomarkers, clinical reasoning regarding disease prediction, multi-modal integration of ophthalmic data and treatment monitoring through time-series analysis. Also, we propose and benchmark medically-grounded contrastive learning strategies that are possible because of the presence of correlated multi-modal data within the introduced dataset. The \texttt{OLIVES} dataset opens new frontiers for training holistic and medically-relevant ML frameworks that mimic the clinical diagnosis pipeline for ophthalmic studies.

%% file: Sections/appendixA.tex
\subsection{Links to Access Dataset}
\label{app: links}
We provide open access to the dataset. The images and labels found in the OLIVES dataset are present at:\\
 \href{https://doi.org/10.5281/zenodo.7105232}{Image Access}\\
 Alternate access to the labels directly can be found at: \\
 \href{https://www.dropbox.com/sh/ehw8pqhekgvka3d/AAA4E1n26pEblK7SSqiiSkpma?dl=0}{Labels Access}\\
The benchmarks provided in the paper are accessible at the following link:  \\
 \href{https://github.com/olivesgatech/OLIVES_Dataset}{Code Access} 

\subsection{Licenses and DOI}
\label{app: license}
The code is associated with an MIT License. The DOI of the dataset is 10.5281/zenodo.6622145. The associated license with the dataset is a Creative Commons International 4 license.

\subsection{Maintenance Plan}
\label{app: maintenance}
The code will be hosted within the github repository specified in Section \ref{app: links}. Instructions and details regarding the dataset will be located at this same repository. Images for the dataset are located at the zenodo directory in Section \ref{app: links}. Labels for these images will be included within this same zenodo dataset after acceptance of the paper. Additional data from other clinical studies will be added over time as part of our partnership with the Retinal Consultants of Texas. Within the Github repository, we will maintain a comprehensive survey of all literature that use the OLIVES dataset. This will include a unified result table and access to publicly available github repositories that benchmark on OLIVES. Furthermore, we anticipate additional applications that make use of the OLIVES dataset and its multi-modal and time-series data (Appendix~\ref{app: apps}) and will update the Github repository with these applications.

\subsection{Dataset Folder Structure}\label{app: structure}
\paragraph{Images} The dataset is split into two folders: Prime and TREX-DME. These correspond to the studies that the respective data originated from. These studies also act as labels for images with diabetic retinopathy (within PRIME folder) and DME (within TREX-DME folder) as these are the disease states studied in their respective trials. Within each clinical study directory there are folders that have the imaging data for each respective patient. Inside of each patient folder is a directory for every visit by each patient. Within every visit folder are folders containing the OCT scans and fundus image for the eye(s) associated with the patient of interest. This structure is consistent in both studies with the only difference being that the TREX DME directory is split into three subdirectories called GILA, Monthly, and TREX that identify specific cohorts of patients. Within every visit, there is a numpy file that is the 3D volume stitched together for the OCT scans of that patient. Additionally, for every patient, there is a numpy file that holds the fundus image and OCT volume generated at every visit into one data structure in the order in which the visits occurred.

\paragraph{Labels}
The labels exist within two directories called "full labels" and "ml centric labels." Full labels contains the complete clinical datasheets for both the Prime and TREX DME studies. This directory also has a word document with additional details regarding the study. The ml centric labels directory has two csv files. The first contains full biomarker and clinical labels for the 9408 OCT scans that were labeled from the first and last visit of every eye.  The other excel file contains the BCVA, CST, eye id, and patient id of all 78185 OCT scans that exist within the OLIVES dataset. These are the clinical labels that are common between both trials. 

\subsection{Reproducibility Statement and Attributions}
\label{app: attributions}
We compare against three self-supervised approaches in this paper. Links to their implementations are provided here:

\href{https://github.com/HobbitLong/SupContrast}{SimCLR } \\
\href{https://github.com/salesforce/PCL} {PCL} \\
\href{https://github.com/facebookresearch/moco}{Moco v2}

Results for our paper can be replicated using the code, images, and labels found in Section \ref{app: links}.

%% file: Sections/appendixB.tex
\subsection{Dataset Comparison}
\input{Tables/DatasetComparisons}

In Table~\ref{tab:dataset_comparison}, we compare OLIVES against existing datasets based on 7 relevant considerations: clinical labels, biomarker labels, time-series data, multi-modal data, disease states, number of images, and number of biomarkers. Among these, biomarker labels and disease state labels have the most semantic overlap and necessitate a clear differentiation with how these are defined. Disease states refer to the overall condition of the eye. For example, an eye can have the overall disease of diabetic retinopathy or any of its variants. However, biomarkers refer to explicit features present within an OCT scan or fundus image that can act as indicators for the disease~\cite{golabbakhsh2013vessel}. For example, a biomarker such as intra-retinal fluid (IRF), is a description of the features present in an individual image, but do not make a statement of the overall disease that the eye is experiencing. Additionally, biomarkers can vary between OCT scans found at different positions within a volume and thus act as a more fine-grained description of the content of an individual image. Furthermore, we define biomarkers with respect to biological features, rather than measurements taken across the image. We deem measurements, such as various retinal thickness values, as a type of clinical label due to its derivation from values taken from the imaging acquisition device (OCT Machine).

\subsection{Challenges in Dataset}
\label{app: challenges}

A number of challenging datasets exist for natural images and videos. These challenges include noise additions~\cite{temel2017cure}, background and imaging modality shifts~\cite{temel2018cure}, fine-grained domain shifted videos~\cite{chen2020action}, and microscopic textures~\cite{hu2021fabric}. Challenging datasets for computed images include large scale seismic datasets~\cite{alaudah2019machine}. The challenge in OLIVES and other medical datasets arises not because of interventions in data, but due to issues in data collection, inversion, representation, annotation, and analysis of minute changes within computed data. Consider Fig.~\ref{fig:challenges}a). A singular OCT scan sampled randomly from the 3D volume of two separate visits between treatments is shown. Notice the same disease diagnosis and minimal differences within the scans. In contrast, Fig.~\ref{fig:challenges}b) shows the OCT scans of three separate patients in their first visit, all of whom are diagnosed with DME. The manifestations of the DME patholology is noticeably different between patients. Similarly, in Fig.~\ref{fig:challenges}c), the CST clinical label for two separate patients with visually dissimlar OCT scans is shown. On the other hand, gradually decreasing CST values between visits for the same patient indicates a decrease in DME's manifestation in Fig.~\ref{fig:challenges}d).

Moreover, the ML techniques used to analyze natural images may not be applicable or sufficient for OCT scans.~\cite{rivail2019modeling} introduced a novel pretext task that involved predicting the time interval between OCT scans taken by the same patient. \cite{zhang2021twin} showed how a combination of different pretext tasks such as rotation prediction and jigsaw re-ordering can improve performance on an OCT anomaly detection task. \cite{qiu2019self} showed how assigning pseudo-labels from the output of a classifier can be used to effectively identify labels that might be erroneous. These works all identify ways to use variants of deep learning to detect important biomarkers in OCT scans. The OLIVES dataset introduces new challenges in these setups by providing biomarkers and clinical labels that correlate with image data.

\subsection{Dataset Logistics}
\paragraph{Clinical Trial Funding} The initial clinical trials, PRIME and TREX-DME are published at~\cite{hannah2021real} and~\cite{payne2017randomized,payne2019randomized,payne2021long,wykoff2019intravitreal} respectively. These trials were conducted between December 2013 and April 2021 at the Retina Consultants of Texas (Houston, TX, USA). The PRIME study was supported by Regeneron Pharmaceuticals. Further financial disclosures are provided in~\cite{hannah2021real}. The corresponding author on~\cite{hannah2021real} is also an author for this article. The TREX-DME study was supported by various grants detailed after References in~\cite{payne2017randomized}.

\paragraph{Labeling} The processes for the clinical trials and diagnosis is provided in ~\cite{hannah2021real} and~\cite{payne2017randomized,payne2019randomized,payne2021long,wykoff2019intravitreal}. For OLIVES, biomarkers are retrospectively added to $9,408$ images. The biomarkers are identified by Charles C. Wykoff with an ophthalmology experience of sixteen years and the labeling is performed by Stephanie Trejo Corona with a grading experience of one year.

\subsection{Addressing Limitations of OLIVES}
\label{app: limitations}
An issue identified in Section~\ref{sec:Conclusion} is that the OLIVES dataset does not provide a global patient distribution. This is a common problem with medical datasets and has sparked research into strategies that can overcome this distributional bias \cite{rieke2020future}. Within the corpus of ophthalmology related studies, there are several datasets that originate from different regions of the world, such as  \cite{abramoff2013automated} from France, \cite{kermany2018labeled} from a collaboration of the USA and China, \cite{li2019attention} from China, and \cite{holm2017dr} from the United Kingdom. It is possible to train with our dataset and test the resulting algorithm with these and others found at \cite{khan2021global} to test for out of distribution performance from cohorts across the world.

Other limitations are addressed in the main paper and relate to the nature of the cohort in our studies. The cohorts chosen are from patients exhibiting some severity level of Diabetic Retinopathy (DR) or Diabetic Macular Edema (DME). As a result, there are no patients that are completely healthy. If it is desirable to guarantee healthy instances within a specific study, then it is possible to augment our dataset with healthy OCT scans or Fundus images from sources such as~\cite{kermany2018identifying}, \cite{abramoff2013automated}, or \cite{farsiu2014quantitative}. 

\begin{figure}[t!]
\centering
\includegraphics[scale = .5]{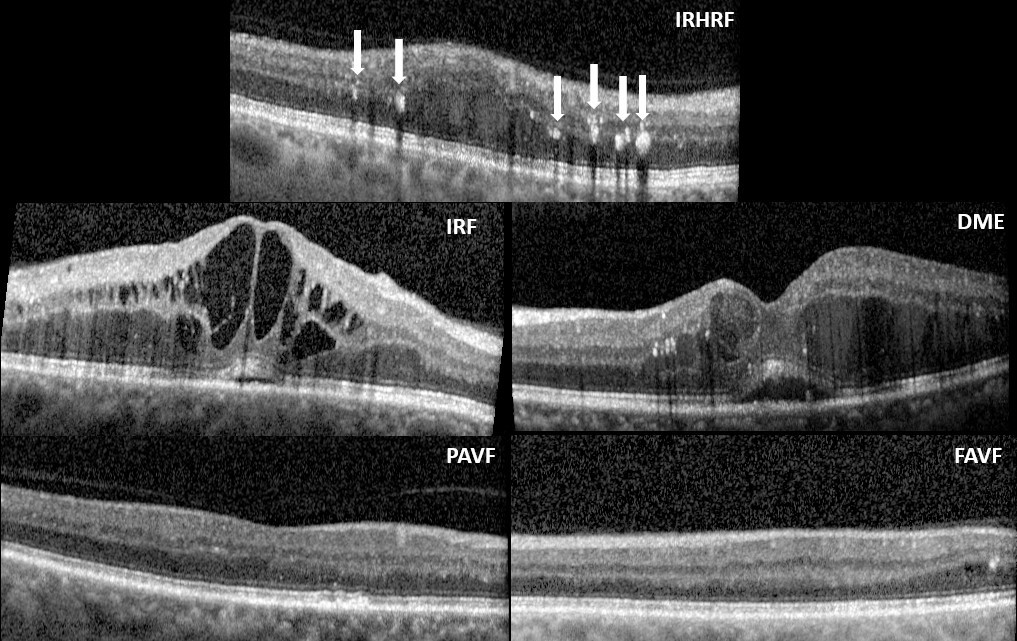}
\label{fig: examples}
\caption{Cross-sectional images of graded biomarkers. Intra-Retinal Hyper-Reflective Foci (IRHRF), indicated by the six white arrows, are areas of hyperreflectivity in the intraretinal layers with or without shadowing of the more posterior retinal layers. Intra-Retinal Fluid (IRF) encompasses the cystic areas of hyporeflectivity. Diabetic Macular Edema (DME) is the apparent swelling and elevation of the macula due to the presence of fluid. A Partially Attached Vitreous Face (PAVF), with an arrow indicating the point of attachment and a Fully Attached Vitreous Face (FAVF). A discussion of these biomarkers can be found at \cite{markan2020novel}.}
\end{figure}

\subsection{Description of Labels}
\subsubsection{Biomarkers and their Generation}
\label{app: biomarkers}
The authors in~\cite{golabbakhsh2013vessel} describe biomarkers as objective indicators of medical state as observed and measured from outside the patient. They are quantifiable characteristics of biological processes. In this paper, the biological processes are diseases and biomarkers indicate the presence or absence of such diseases. Under limited circumstances, the authors in~\cite{golabbakhsh2013vessel} suggest that biomarkers can be surrogate endpoints in clinical trials. However, they caution against doing so unless the underlying clinical trial is specifically meant for the study. As such, biomarkers indicate the presence of diseases, but are not causal to these diseases. Causality in the medical domain can be singular causality or general causality~\cite{rizzi1992causality}. Singular causality is constrained by events in a time-series linked events while general causality analyzes relationships between events. As such this is different from visual causal features from~\cite{prabhushankar2021extracting} or causal question-based analysis in~\cite{alregib2022explanatory} or causal factor analysis in~\cite{chalupka2014visual}.

\input{Tables/Abbreviations}

All image interpretations were performed by a trained grader for the presence of the following parameters: atrophy or thinning of retinal layers, disruption of the ellipsoid zone (EZ), disruption of the retinal inner layers (DRIL), intraretinal (IR) hemorrhages, intraretinal hyperreflective foci (IRHRF), partially attached vitreous face (PAVF), fully attached vitreous face (FAVF), preretinal tissue or hemorrhage, vitreous debris, vitreomacular traction (VMT), diffuse retinal thickening or macular edema (DRT/ME), intraretinal fluid (IRF), subretinal fluid (SRF), disruption of the retinal pigment epithelium (RPE), serous pigment epithelial detachment (PED), and subretinal hyperreflective material (SHRM). The following describes the grading used for each morphological feature evaluated in each B-scan using the Heidelberg Spectralis HRA+OCT software. 

Atrophy or thinning of retinal layers was indicated as present with evidence of RPE atrophy or thinning of the retina at the trained grader’s discretion \cite{mazumder2017spectropathology, xu2011rpe}. Disruption of the EZ was indicated as present with when the second-most posterior hyperreflective band of the retina was discontinuous. DRIL was indicated as present when the boundaries of the retinal inner layers such as the inner nuclear layer, outer plexiform layer, and ganglion cell layer were not clearly defined \cite{sun2014disorganization}.  Intraretinal hemorrhages were indicated as present when there was a small, localized lesion that caused shadowing of the more posterior retinal layers, with a corresponding lesion visible on the near-infrared fundus image. IRHRF were indicated as present with the appearance of intraretinal, highly reflective spots, which correspond pathologically to microaneurysms or hard exudates, with or without shadowing of the more posterior retinal layers \cite{gella2014spectral}. A partially attached vitreous face was indicated as present with evidence of perifoveal detachment of the vitreous from the internal limiting membrane (ILM) with a macular attachment point within a 3-mm radius of the fovea. A fully attached vitreous was indicated as present with no evidence of perifoveal or macular detachment from the ILM. Preretinal tissue or hemorrhage was indicated as present with evidence of an hyporeflective preretinal tissue, epiretinal membrane, or hemorrhage over the surface of the ILM \cite{itoh2016prevalence}. Vitreous debris was indicated as present with evidence of hyperreflective foci in the vitreous or shadowing of the retinal layers in the absence of an intraretinal hemorrhage. VMT was indicated as present with evidence of perifoveal vitreous separation, vitreomacular attachment, and foveal anatomic distortions \cite{duker2013international}. Diffuse retinal thickening or macular edema was indicated as present when there was increased retinal thickness of 50 µm above the otherwise flat retina surface with associated reduced reflectivity in the intraretinal tissues \cite{trichonas2014optical}. Intraretinal fluid was indicated as present when intraretinal hyporeflective areas or cysts had a minimum fluid height of 20 µm \cite{trichonas2014optical}. Subretinal fluid was indicated as present when hyporeflective areas or cysts were evident in the subretinal space between the EZ and RPE layers. Disruption of the RPE was indicated as present when the most posterior hyperreflective band of the retina was discontinuous. Serous pigment epithelial detachment was indicated as present with evidence of a hyporeflective area underneath the detached RPE. SHRM was indicated as present when hyperreflective foci were evident in the subretinal space between the EZ and RPE layers.

\input{Tables/Table_Full}

\subsubsection{Clinical Labels and their Generation}
\paragraph{Full Clinical Labels}\label{app: all_labels}
The clinical labels obtained from the PRIME trials include BCVA, DRSS, CST, eye ID, patient ID, diabetes type, BMI, age, race, gender, HbA1c, leakage index, years with diabetes, and injection arm. The clinical labels from the TREX-DME trials include BCVA, Snellen score, CST, Eye ID, and Patient ID. Since OLIVES is a combination of the two, we use only the common labels from both trials as our clinical labels in our experiments. These common labels include BCVA, CST, Patient ID and Eye ID which are listed in Table~\ref{tab:olives}. However, we provide access to all available labels as described in Appendix~\ref{app: structure}.

The Early Treatment Diabetic Retinopathy Study (ETDRS) diabetic retinopathy severity scale (DRSS) has 13 levels describing DR severity and change over time based on color fundus photograph grading. The scale starts at level 10 and ends at level 90 with irregular scale numbering. Nonproliferative diabetic retinopathy (NPDR) DRSS levels on the scale are below 61 and proliferative diabetic retinopathy (PDR) levels are 61 and above. Diabetes type refers to the patient's diagnosis of either type one or type two diabetes mellitus. HbA1c is the measurement of glycated hemoglobin, commonly referred to as blood sugar, which serves as an indicator for diabetes diagnosis or diabetic control. Leakage index refers to the panretinal leakage index used in the PRIME trial in which areas of leakage, regions of hyperfluorescence in fluorescein angiography images, were divided by areas of interest, region of total analyzable retinal area, and converted to a percentage. Injection arm refers to either the DRSS-guided (1) cohort or the PLI-guided (2) cohort in the PRIME trial. Snellen score is the visual acuity testing procedure commonly used in ophthalmic clinical settings. The first number indicates the distance in feet that the letter chart was read, in U.S., this number is commonly 20, followed by a number indicating the distance a person with "normal" vision (20/20) would have to be to read something the person tested could read at 20 feet. Thus, a larger denominator would indicate poorer vision.

Other self-explanatory demographic information including body mass index (BMI), age, race, and gender are provided. We caution the users regarding the societal impact of using these labels since the underlying PRIME trial did not study the causality of these labels.

\paragraph{ML Centric Clinical Labels}\label{app: labels}
We describe BCVA and CST in this section. ETDRS best-corrected visual acuity (BCVA) is a visual function assessment performed by certified examiners where a standard vision chart is placed 4-meters away from the patient. The patient is instructed to read the chart from left to right from top to bottom until the subject completes 6 rows of letters or the subject is unable to read any more letters. The examiner marks how many letters were correctly identified by the patient. Central subfield thickness (CST) is the average macular thickness in the central 1-mm radius of the ETDRS grid. CST was obtained from the automated macular topographic information in the Heidelberg Eye Explorer OCT software.

The remaining clinical labels of Patient ID and Eye ID are self-explanatory and collected on clinical visits. 

\begin{figure}[t]
\centering
\includegraphics[scale = .6]{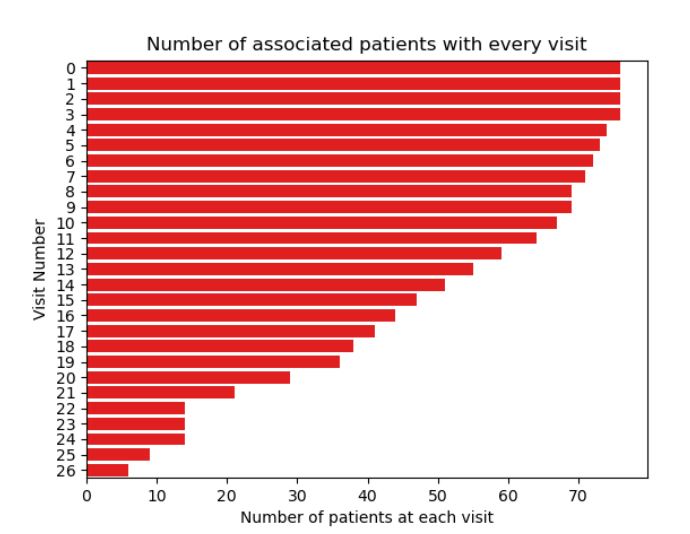}
\vspace{-1.5mm}
\caption{Number of patients at every visit within one of the training sets used for the treatment predicition analysis.}\vspace{-1.5mm}\label{fig: patient_dist}
\end{figure}

\begin{figure}[h!]
\centering
\includegraphics[scale = .55]{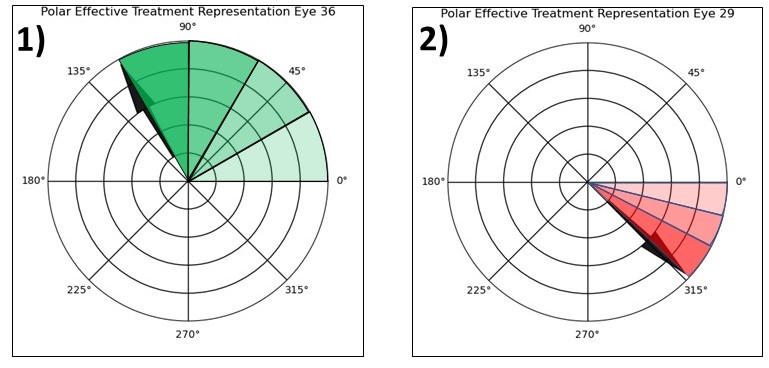}
\caption{1) A plot of average number of visits by patients that were an improvement or deterioration from previous week. Red bars indicate the standard deviation across all patients. 2) Plot of average change in BCVA with respect to the first week. 1) and 2) Polar representation plots with respect to the effective treatment of individual patients. The angle of rotation is $\theta$ = 360 / (number of visits for patient). The angle rotates by $\theta$ counterclockwise if the collected BCVA on the current visit is better than that of the previous visit and clockwise by $\theta$ if it's worse.}\label{fig: progression_app}
\end{figure}

\subsubsection{Time Series Labels}
\label{app: time_series_labels}

The labels generated for the time-series experiments were based on changes in week to week BCVA values. For an individual visit, the treatment label was set to 1 if the following visit resulted in an improvement in BCVA and a 0 if the result wasn't an improvement. The goal is to predict whether the next visit would result in an improvement based on the associated modality (Fundus or 3D OCT Volume). Fig.~\ref{fig: progression} provides statistics regarding visit-wise changes of BCVA within the dataset. Further analysis of the dataset requires the number of patients treated on each visit which is provided in Fig.~\ref{fig: patient_dist}. As is apparent, the number of patients keep decreasing across visits. This can be for a variety of reasons all of which are discussed in the clinical trail publications at~\cite{hannah2021real} and~\cite{payne2017randomized,payne2019randomized,payne2021long,wykoff2019intravitreal}. These numbers provide further context to the changes in Fig.~\ref{fig: progression}. Presumably, as the treatment continues, it is the challenging patients who return for treatment and who qualify for injections. Their visit-wise average BCVA change skews the cohort in the negative direction in Fig.~\ref{fig: progression}. 

This change in treatment improvements can be understood through the polar plots of Figure \ref{fig: progression_app}. The polar representation plots show the effective treatment with respect to an individual patient. This works by a vector beginning at the 0 degree point and rotating by an angle $\theta$ that is 360/(number of patient visits). After each turn counterclockwise, the hue of the associated color becomes darker by a fixed degree with a darker green hue indicating a higher degree of improvement and a darker red hue indicating a higher degree of worsening in the clockwise direction. This is shown in plots 1 and 2 in Fig.~\ref{fig: progression_app}. Plot 1 shows that the patient had 4 rotations as indicated by the hue of green becoming darker by four degrees. This indicates that out of the total number of visits, this patient experienced 4 more visits with improvements, rather than deterioration. The converse is true for Plot 2. 

%% file: Tables/DatasetComparisons.tex
\begin{table}[h!]
\small
\centering
\begin{tabular}{@{}cccccccc@{}}
\toprule
% \multicolumn{8}{c}{Ophthalmology Dataset Comparison}   
% \\ \midrule
\multicolumn{1}{c}{Dataset} &  \multicolumn{1}{c}{Clinical} & \multicolumn{1}{c}{Biomarker} &
\multicolumn{1}{c}{TimeSeries} & 
\multicolumn{1}{c}{MultiModal} & 
\multicolumn{1}{c}{Disease} &

\multicolumn{1}{c}{No. of} &
\multicolumn{1}{c}{No. of} \\
&  \multicolumn{1}{c}{Labels} & \multicolumn{1}{c}{Labels} &
\multicolumn{1}{c}{Data} & 
\multicolumn{1}{c}{Images} & 
\multicolumn{1}{c}{States} &
\multicolumn{1}{c}{Images} & 
\multicolumn{1}{c}{Biomarkers}\\
\hline
\midrule
Kermany \cite{kermany2018labeled} & \checkmark & \checkmark & \xmark & \xmark & \checkmark & 109312 & 4\\ 
Farisu \cite{farsiu2014quantitative} & \checkmark & \checkmark & \xmark & \xmark & \checkmark & 38400 & 4 \\
Srinivasan \cite{srinivasan2014fully} & \checkmark & \xmark & \xmark & \xmark & \checkmark & 3231 & 0 \\
Maetschke \cite{maetschke2019feature} & \checkmark & \xmark & \xmark & \xmark & \checkmark & 1110 & 0\\\midrule 

Kaggle DR \cite{kaggle} & \checkmark & \xmark & \xmark & \xmark & \checkmark & 35126 & 0 \\ 
AG-CNN \cite{li2019attention} & \checkmark & \xmark & \xmark & \xmark & \checkmark & 4854 & 0 \\ 
ODIR \cite{li2020benchmark} & \checkmark & \xmark & \xmark & \xmark & \checkmark & 10000 & 0 \\ 
DeepDrid \cite{liu2022deepdrid} & \checkmark & \xmark & \xmark & \checkmark & \checkmark & 2256 & 0\\ 
Laterality \cite{liu2019self} & \checkmark & \xmark & \xmark & \xmark & \checkmark & 18394 & 0\\ 
Messidor \cite{abramoff2013automated} & \checkmark & \xmark & \xmark & \xmark  & \checkmark & 1748 & 0\\ \midrule

OLIVES & \checkmark & \checkmark & \checkmark & \checkmark & \checkmark & 78185 & 16\\
\bottomrule
\end{tabular}
\caption{Comparison of eye-related datasets along relevant medical considerations.}
\label{tab:dataset_comparison}
\end{table}

\begin{table}[h!]
\small
\centering
\begin{tabular}{@{}cccccccc@{}}
\toprule
% \multicolumn{8}{c}{Ophthalmology Dataset Comparison}   
% \\ \midrule
\multicolumn{1}{c}{Dataset} &  \multicolumn{1}{c}{Clinical} & \multicolumn{1}{c}{Biomarker} &
\multicolumn{1}{c}{MultiModal} & 
\multicolumn{1}{c}{Disease} &
\multicolumn{1}{c}{No. of} &

\multicolumn{1}{c}{No. of} &
\multicolumn{1}{c}{No. of} \\
&  \multicolumn{1}{c}{Labels} & \multicolumn{1}{c}{Labels} &
\multicolumn{1}{c}{Images} & 
\multicolumn{1}{c}{States} &
\multicolumn{1}{c}{Eyes} & 

\multicolumn{1}{c}{Images} & 
\multicolumn{1}{c}{Biomarkers}\\
\hline
\midrule
Rotterdam \cite{adal2015accuracy} & \checkmark & \xmark & \xmark & \checkmark & 70 & 1120 & 0 \\
Rivail et. al.~\cite{rivail2019modeling} & \xmark & \xmark & \xmark & \checkmark & 221 & 3308 & 0\\
OLIVES & \checkmark & \checkmark & \checkmark & \checkmark & 96 & 78185 & 16\\
\bottomrule
\end{tabular}
\caption{Comparison of eye-related time-series datasets along relevant medical considerations.}
\label{tab:dataset_comparison_time}
\end{table}

%% file: Tables/Abbreviations.tex
\begin{table}[]
\centering
\begin{tabular}{@{}cc@{}}
\toprule

\multicolumn{2}{c}{Table of Abbreviations}              \\ \midrule
\multicolumn{1}{c}{Abbreviation} & \multicolumn{1}{c}{Full Name}              \\ \midrule
CST                                & Central Subfield Thickness                  \\
BCVA                               & Best Central Visual Acuity                  \\
Eye ID                             & Eye Identity                                \\
EZ                                 & Ellipsoid Zone                              \\
DRIL                               & Disruptionof the Retinal Inner Layers       \\
IR                                 & Intraretinal                                \\
IRHRF                              & Intraretinal Hyperreflective Foci           \\
PAVF                               & Partially Attached Vitreous Face            \\
FAVF                               & Fully Attached Vitreous Face                \\
VMT                                & Vitreomascular Traction                     \\
DRT/ME                             & Diffuse Retinal Thickening or Macular Edema \\
IRF                                & Intraretinal Fluid                          \\
SRF                                & Subretinal Fluid                            \\
RPE                                & Retinal Pigment Epithelium                  \\
PED                                & Pigment Epithelial Detachment               \\
SHRM                               & Subretinal Hyperreflective Material         \\
DR                                 & Diabetic Retinopathy                        \\
DME                                & Diabetic Macular Edema                      \\
CI-DME                                &  Center-Involved Diabetic Macular Edema              \\
PDR                                &   Proliferative Diabetic Retinopathy              \\
NPDR                                & Non-Proliferative Diabetic Retinopathy                \\

OCT                                & Optical Coherence Tomography                \\
AMD                                &    Age-related Macular Degeneration             \\
CNV                                &  Choroidal Neovascularization               \\
VEGF                        &    Vascular Endothelial Growth Factor                   \\
ETDRS                   &   Early Treatment Diabetic Retinopathy Study              \\
DRSS                    & Diabetic Retinopathy Severity Scale \\
PRIME                   & Real-Time Objective Imaging to Achieve Diabetic Retinopathy Improvement\\
TREX-DME                & Treat and Extend Protocol in Patients with Diabetic Macular Edema\\

\bottomrule
\end{tabular}\vspace{+1mm}\caption{Summary clinical and biomarker abbreviations used throughout the paper.}
\label{tab: abbreviation}
\end{table}

%% file: Tables/Table_Full.tex
\begin{table*}[]
\tiny
\begin{tabular}{@{}lllll@{}}
\toprule
\multicolumn{1}{c}{Dataset} &
  \multicolumn{1}{c}{Statistics} &
  \multicolumn{1}{c}{Label Type} &
  \multicolumn{2}{c}{Label Names} \\ \hline \midrule
\multicolumn{1}{l}{PRIME Clinical} &
  \multicolumn{1}{l}{\begin{tabular}[c]{@{}l@{}}29000+ Images\\ 40 Patients\\ 40 Unique Eyes\end{tabular}} &
  \multicolumn{1}{l}{Clinical} &
  \multicolumn{2}{l}{\begin{tabular}[c]{@{}c@{}}BCVA, CST, DRSS, Eye ID,Patient ID, Diabetes Type, BMI, Age, Race, Gender \\ Years with Diabetes, HbA1c, Leakage Index, Injection Arm \\\end{tabular}} \\ \midrule
\multicolumn{1}{l}{\multirow{2.5}{*}{PRIME Biomarker}} &
  \multicolumn{1}{l}{\multirow{2}{*}{\begin{tabular}[c]{@{}l@{}}3900+ Images\\ 40 Patients\\ 40 Unique Eyes\end{tabular}}} &
  \multicolumn{1}{l}{Clinical} &
  \multicolumn{2}{l}{\begin{tabular}[c]{@{}c@{}}BCVA, CST, DRSS, Eye ID,Patient ID, Diabetes Type, BMI, Age, Race, Gender \\ Years with Diabetes, HbA1c, Leakage Index, Injection Arm \\\end{tabular} } \\ [.5 pt] \cmidrule(l){3-5} 
\multicolumn{1}{l}{} &
  \multicolumn{1}{l}{} &
  \multicolumn{1}{l}{Biomarker} &
  \multicolumn{2}{l}{16 Biomarkers (DME, IRF, IRHRF, etc.)} \\ \midrule
\multicolumn{1}{l}{TREX-DME Clinical} &
  \multicolumn{1}{l}{\begin{tabular}[c]{@{}l@{}}38000+ Images\\ 47 Patients\\ 56 Unique Eyes\end{tabular}} &
  \multicolumn{1}{l}{Clinical} &
  \multicolumn{2}{l}{BCVA, Snellen Score, CST, Eye ID, Patient ID} \\ \midrule
\multicolumn{1}{l}{\multirow{2.5}{*}{TREX-DME Biomarker}} &
  \multicolumn{1}{l}{\multirow{1}{*}{\begin{tabular}[c]{@{}l@{}}5300+ Images\\ 47 Patients\\ 56 Unique Eyes\end{tabular}}} &
  \multicolumn{1}{l}{Clinical} &
  \multicolumn{2}{l}{BCVA, Snellen Score, CST, Eye ID, Patient ID} \\ \cmidrule(l){3-5} 
\multicolumn{1}{l}{} &
  \multicolumn{1}{l}{} &
  \multicolumn{1}{l}{Biomarker} &
  \multicolumn{2}{l}{16 Biomarkers (DME, IRF, IRHRF, etc.)} \\ \midrule
\multicolumn{1}{l}{\multirow{2.5}{*}{TREX-DME + PRIME Biomarker}} &
  \multicolumn{1}{l}{\multirow{2}{*}{\begin{tabular}[c]{@{}l@{}}9200+ Images\\ 87 Patients\\ 96 Unique Eyes\end{tabular}}} &
  \multicolumn{1}{l}{Clinical} &
  \multicolumn{2}{l}{BCVA, CST, Eye ID, Patient ID} \\ \cmidrule(l){3-5} 
\multicolumn{1}{l}{} &
  \multicolumn{1}{l}{} &
  \multicolumn{1}{l}{Biomarker} &
  \multicolumn{2}{l}{16 Biomarkers (DME, IRF, IRHRF, etc.)} \\ \midrule
\multicolumn{1}{l}{\multirow{1}{*}{TREX-DME + PRIME Clinical}} &
  \begin{tabular}[c]{@{}l@{}}67000+ Images\\ 87 Patients\\ 96 Unique Eyes\end{tabular} &
   \multicolumn{1}{l}{Clinical} &
   \multicolumn{2}{l}{BCVA, CST, Eye ID, Patient ID} \\ \bottomrule
\end{tabular}
\caption{Summary of clinical and biomarker data present within each individual study.}
\label{tab: data_all}
\end{table*}

%% file: Sections/appendixC.tex
\subsection{Multi-Modal Integration Between OCT and Biomarker/Clinical Labels}

\begin{table}[h!]  
\small
  \begin{center}
    \caption{Benchmark results for DR/DME detection showing precision and recall.}
    \label{tab:detection benchmarks appendix}
    \begin{tabular}{*{16}{c c c c c c}}
      \toprule % <-- Toprule here
      \multirow{2}{*}{\textbf{Experiments}}  &
      \multirow{2}{*}{\textbf{Model}}  &
    %   \multirow{2}{*}{\textbf{Balanced Accuracy}} &
      \multicolumn{2}{c}{\textbf{Precision}} 
        & \multicolumn{2}{c}{\textbf{Recall}} \\
      & & DR & DME & DR & DME \\
      \hline
      \midrule % <-- Midrule here
      OCT & R-18 & 0.747 & 0.670 & 0.608 & 0.794 \\
      Clinical & MLP & 0.753 & 0.756 & 0.758 & 0.751 \\
      Biomarker & MLP & 0.703 & 0.870 & 0.826 & 0.771 \\
      OCT + Clinical & R-18 + MLP & 0.888 & 0.765 & 0.566 & 0.952 \\
      OCT + Biomarker & R-18 + MLP & 0.885 & 0.778 & 0.742 & 0.904 \\
      \bottomrule % <-- Bottomrule here
    \end{tabular}
  \end{center}
\end{table}

\label{app: multi-modal details}
\paragraph{Experimental Details}
The greyscale B-scans are rescaled to $128 \times 128$ and normalized with $\mu=0.482$ and $\sigma=0.037$ as the baseline data for DR/DME detection. For OCT, we utilize Resnet-18 (R-18) \cite{he2016deep} along with Adam optimizer and a learning rate of $1.5e-4$. There are 20 eyes in the test set; 10 having DR and the remaining DME. The validation set has 5 eyes with DR and the other 5 exhibiting DME. The train set is composed of the remaining 66 eyes, 26 of which have DR while 40 have DME. Therefore, we utilize $6, 468$ images in the training set, $1,960$ images in our test set and $980$ images in the validation set. For supervised learning with clinical labels, we train a shallow Multi Layer Perceptron (MLP) with two linear layers and Relu activation between. Biomarker features are normalized to zero mean and unit standard deviation. For supervised learning with biomarkers, we train a shallow MLP with four linear layers and LeakyRelu activation between. Biomarker features are normalized to zero mean and unit standard deviation. For multi-modal learning with OCT and clinical labels/biomarkes, we use the same train, test and validation split as the baseline OCT model and the clinical labels/biomarkers associated with each B-scan.

\label{app: guided loss}
\paragraph{Optimization via Guided Loss}
Each modality is input to its independent model. At the output of the MLP biomarkers/clinical label model are logits $\phi^{MLP}(x_i)$, while the output logits of the Resnet OCT model are $\phi^{Resnet}(x_i)$. During optimization, learned features from one modality (biomarkers or clinical labels) are used to optimize the learning of the other (OCT features). The guided loss, $\mathcal{L}_{Guided}$, is one component of the overall loss function $\mathcal{L}$. Guided loss is the mean square error between MLP logits and Resnet logits. At every epoch, we minimize the disparity between these logits until the stopping criteria for training is met. The other two components, $\mathcal{L}_{Resnet}$ and $\mathcal{L}_{MLP}$ are binary cross entropy losses computed between the ground truth labels and logits from each model respectively. Collectively the three terms allow a joint optimization of both models and a transfer of knowledge from MLP model to Resnet model. 

\begin{align}
\mathcal{L} = \mathcal{L}_{Resnet} + \mathcal{L}_{MLP} + \mathcal{L}_{Guided}
\label{eq:pct}
\end{align}

\begin{align}
    \mathcal{L}_{Guided} = \textbf{1} \left[ \hat{y}^{MLP} = y \right] \frac{1}{2} \left|\left|\phi^{Resnet}(x_i) - \phi^{MLP}(x_i) \right|\right|_2^2 
\end{align}

\paragraph{Medical Perspective of Benchmark Results}
The results show that biomarkers as features are more effective at discriminating between the disease classes in both uni- and multi-modal training scenarios. This makes sense from a medical perspective because biomarker features have direct correlation to the presence of DR/DME. Also, biomarker vectors assigned to any OCT slice are specific features that visually manifest themselves within that OCT slice. This means that biomarker features are fine-grained signs of diagnostic patterns indicative of disease. Clinical labels on the other hand are more coarse. Some clinical labels, like CST, represent characteristics of an OCT volume as a whole rather than any specific slice within a volume. Other clinical labels, like BCVA, are not derived from OCT and represent an evaluation of the eye as a whole. CST and BCVA are clinical parameters that are not indicative of a specific retinal disease diagnosis, but are instead representations of retinal anatomy or visual function, respectively. These two features are used in the context of monitoring retinal disease progression; thus, it is unsurprising that within a machine learning framework they yield sub-optimal performance in discriminating between disease classes.

\subsection{Biomarker Interpretation with Contrastive Learning}
\label{app: sup_con}
A number of variations of contrastive learning exist in literature. The authors in~\cite{prabhushankar2020contrastive} use the term contrastive to design visual explanations. They then extend these explanations to perform contrastive reasoning in inferential framework in~\cite{prabhushankar2021contrastive}.

This clinically aware supervised contrastive loss can be represented by: 
\begin{align}
    L_{supcon_{clinical}} = \sum_{i\in{I}} \frac{-1}{|C(i)|}\sum_{c\in{C(i)}}log\frac{exp(z_{i}\cdot z_{c}/\tau)}{\sum_{a\in{A(i)}}exp(z_{i}\cdot z_{a}/\tau)}
\end{align}

where $i$ is the index for the image of interest $x_{i}$.
All positives $c$ for image $x_{i}$ are obtained from the set $C(i)$ and all positive and negative instances $a$ are obtained from the set $A(i)$. Every element $c$ of $C(i)$ represents all other images in the batch with the same clinical label $c$ as the image of interest $x_{i}$. Additionally, $z_{i}$ is the embedding for the image of interest, $z_{c}$ represents the embedding for the clinical positives, and $z_{a}$ represents the embeddings for all positive and negative instances in the set $A(i)$. Embeddings are obtained after passing the representations from an encoder network $f(.)$ through a projection head $G(.)$ that we set to be a multi-layer perceptron network. $\tau$ is a temperature scaling parameter that is set to .07 for all experiments. For example, a loss represented as $L_{BCVA}$ indicates a supervised contrastive loss where BCVA is utilized as the clinical label of interest and all positives are chosen based on having the same BCVA value as the target image.

\paragraph{Training} In this study, we leverage knowledge learnt from training on the large set of clinical labels to improve performance in classifying the smaller set of biomarkers. To test this setup, we take 76 eyes from the \texttt{OLIVES} Dataset to form a training set and take the remaining set of 20 eyes to form a test set. From this set of 20 eyes, we form an individual balanced test set for each biomarker by sampling 500 OCT scans with the biomarker present and 500 OCT scans with the biomarker absent. We train for 25 epochs and utilize a stochastic gradient descent optimizer with a learning rate of $1e-3$ and momentum of $0.9$. The applied augmentations are random resize crops of size of 224, random horizontal flips, random color jitter, and data normalization to the mean and standard deviation of the respective dataset with a batch size set at 64. We use Intraretinal Hyperreflective Foci (IRHRF), Partially Attached Vitreous Face (PAVF), Fully Attached Vitreous Face (FAVF), Intraretinal Fluid (IRF), and Diffuse Retinal Thickening or Diabetic Macular Edema (DRT/ME) as the biomarkers in the study. 

\paragraph{Medical Interpretation of Benchmark Results}
Another aspect of the results is how well the used clinical labels correspond with the biomarker classification performance. In all cases, the results act as validation to the hypothesis that taking advantage of correlations that exist with certain clinical labels is beneficial for biomarker detection of individual OCT scans. However, from a medical perspective, certain outcomes would intuitively be more likely. For example, the severity of IRF and DME tend to be correlated with CST due to higher levels of fluid corresponding to changes in CST. Therefore, it makes sense that the best performance for IRF and DME is associated with using CST values as the clinical label for the loss. Additionally, it can be observed in Fig.~\ref{fig: distribution} that because BCVA and CST have different distributions of values,  there is a different number of associated eyes and images for each respective clinical value. Effectively, this means that there is varying diversity with respect to any individual clinical label, which explains the varying performance depending on which clinical label is used. The Eye ID works due to images from the same eye having many features in common that serve to identify a good positive set for the loss. However, from a medical perspective, the Eye ID alone does not confer any additional medical insight. 

\input{Tables/ContrastError}

\subsection{Time-series Treatment Analysis}
\paragraph{Experimental Procedure for Predicting Successive Treatment Effect} 
\label{subsec:treatmenteffect}
To perform this experiment, we generate treatment effect labels. For every OCT volume or fundus image, we assign a label $1$ if the following visit resulted in an increase in BCVA and a label $0$ if the next week resulted in a decrease in BCVA. We then train models to perform this binary classification task of next visit improvement or deterioration. Each architecture is trained for 25 epochs with a SGD optimizer, learning rate of .0001, momentum of .9, and a batch size of 10. We use a Resnet-18 \cite{he2016deep}, ResNet-50 \cite{he2016deep}, DenseNet-121 \cite{huang2017densely}, EfficientNet \cite{tan2019efficientnet}, and Vision Transformer~\cite{dosovitskiy2010image} (using a patch size of 32, 16 transformer blocks, 16 heads in multi-attention layer). EfficientNet on OCT performs with the best results from Table~\ref{tab:bcva_pred}. It can also be observed that the vision transformer model did not significantly improve over the traditionally CNN models. It is possible that the attention mechanism of the transformer needs further training and refinement to learn patches of importance  within a Fundus image. This is especially true within the context of medical data as small fine-grained locations are oftentimes the most important and difficult to identify. From the overall results, it is clear that the main bottleneck to good performance is overfitting of the model towards a single class, which, in this case, was the treatment effect. This makes sense as these volumes tend to have more readily distinguishable features due to a more severe variation of the disease. Also, the best performance in Table \ref{tab:bcva_pred} came when using OCT Volumes as well as the smaller ResNet-18 and EfficientNet models which may be due to having more data than in the Fundus case as well as less prone to overfitting due to a smaller size.

\paragraph{Medical Perspective of Predicting Successive Treatment Effect}

In Table~\ref{tab:bcva_pred}, it is observed that the model is able to predict whether the next visit will experience an improvement or worsening of BCVA on the following visit with the associated performance seen in this table. From a medical perspective, indicators that predict whether treatment will be successful or not is not so clear simply from imaging data. This is reflected by performance measures for accuracy that are barely better than random chance. Part of the challenge is that responses to the treatment could potentially be due to factors independent from the imaging data. For example, lifestyle choices on the part of the patient could have a corresponding impact on how well the treatment is able to perform. Additionally, patients do not receive treatments equally due to the specific nature of an individual's condition, which limits predictability. In order to improve upon this benchmark, future studies should investigate the effect of utilizing more powerful time-series models as well as multi-modal fusion of fundus, OCT, clinical, and biomarker data.

\label{app: time-series}
\begin{figure}[h!]
\centering
\includegraphics[scale = .35]{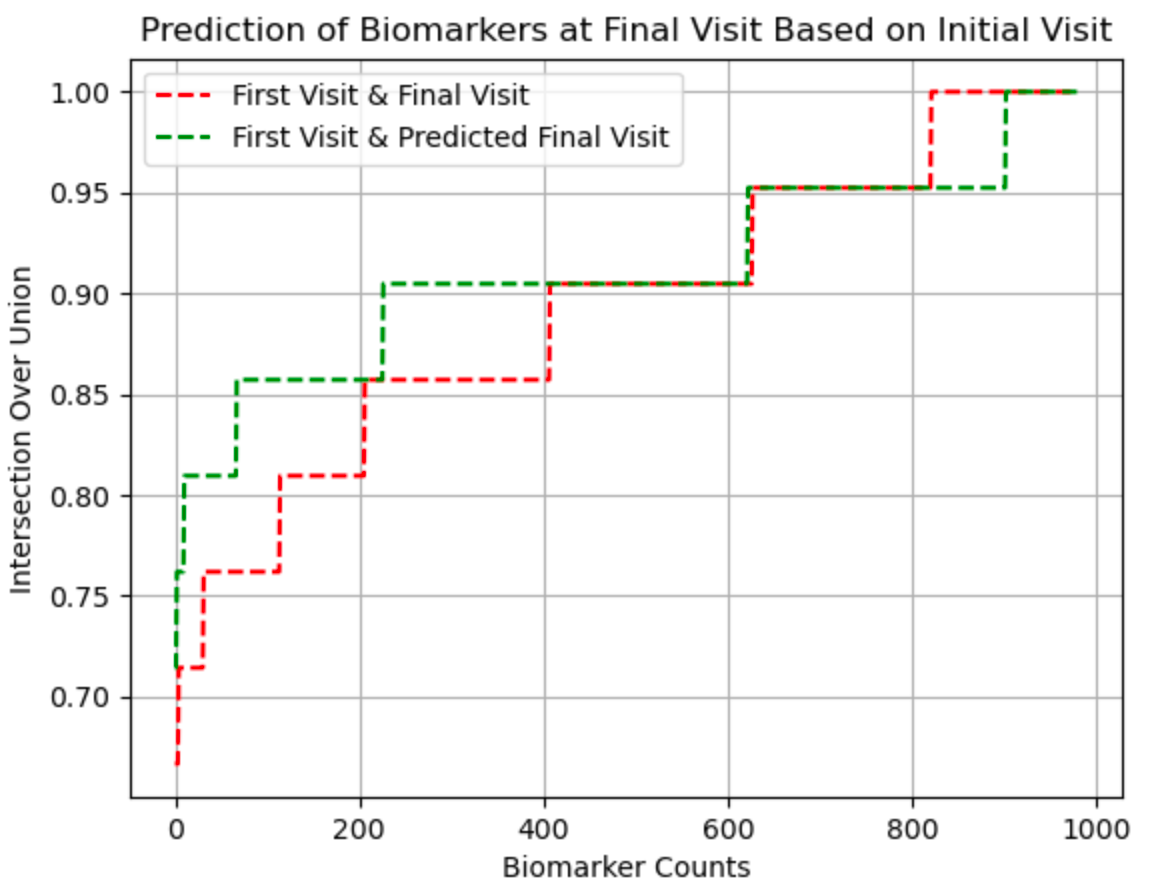}
\caption{Benchmark Prediction of final visit biomarkers from initial visit biomarkers}\label{fig:final_bio_pred}
\end{figure}

\paragraph{Predicting Final Ocular State}
We perform a similar analysis with the biomarkers available from patient's first and final visit. In this analysis we explore the predictive power of the initial biomarkers at forecasting the biomarkers at the final visit. $3,234$ biomarkers were used as input features in the training set while $980$ and $490$  biomarkers were used for test and validation set respectively. A shallow MLP of two linear layers and a Relu activation was the model used in this analysis. Intersection over union served as the metric to evaluate the quality of the predicted final visit biomarkers relative to the ground truth. Figure \ref{fig:final_bio_pred}, shown in Appendix \ref{app: time-series}, shows the overall performance of the model on the test set. We see that initial biomarkers serve as good features for prediction of final ocular state.

\paragraph{Medical Perspective of Predicting Final Ocular State} 
This axes of Fig.~\ref{fig:final_bio_pred}, show the number of biomarker vectors in the test set having varying intersection over union values. The red curve is the intersection over union (IOU) between the biomarkers at the first and final visits. This is the reference to show how final visit biomarkers changed relative to initial biomarkers. From the red curve, we see that of the 980 samples in the test set, only a few, approximately $200$, have an IOU of 1. This means these biomarkers remained constant between first and final visits. A final visit biomarker being the same as the initial may not be an indicator of the patient's response to treatment. Rather, it indicates that no additional biomarkers manifested themselves at the final visit. The green curve represents the IOU between initial visit biomarkers and the predicted final visit biomarkers. Ideally, a complete overlap of red and green curves is desired. This would indicate that changes in biomarkers between first and final visits are being properly captured by the model. Complete overlap occurred for approximately 400 biomarkers whose IOU range from $0.90 - 0.95$. The time when the green curve exceeds the red indicates when there are larger intersection between first and predicted final visit biomarkers compared to the intersection between first and true final visit biomarkers. This means that the model predicted additional biomarkers within those biomarker vectors than what is actually present at the final visit. Conversely, there are a few cases when the green curve recedes the red and these are times when the model predicted fewer biomarkers within the biomarker vectors than the actual amount present at the final visit.

\subsection{Other potential applications}
\label{app: apps}
The rich set of labels in the OLIVES dataset allows for utilizing the clinical labels and biomarkers in multiple ways. We demonstrate multi-modal fusion, medically-grounded contrastive learning, and time-series predictions in Section~\ref{sec:Results}. In addition, Active Learning~\cite{logan2022decal, logan2022patient} can utilize the clinical labels and biomarkers as indicators of disease states. The two paradigms of active learning - uncertainty and diversity - can be derived not through the model predictions, but from the auxiliary data in OLIVES. Similarly, biomarkers provide an annotated set of visual characteristics that show the manifestations of diseases within OCT scans. These biomarkers, along with the disease states and OCT scans, can be utilized for clinical reasoning. Other potential applications include domain difference analysis and adaptation which is described in Sections.~\ref{app: domain-Shift} and~\ref{app: domain adaptation} respectively.

\subsection{Domain Shift}\label{app: domain-Shift}
\input{Tables/Domain_Shift_Trial}
\input{Tables/Domain_Shift_Treatment}

\paragraph{Domain shift based on PRIME and TREX trials} Both the PRIME and TREX DME clinical trials are conducted using the same imaging equipment, in the same clinic. Hence, the domain difference w.r.t. PRIME and TREX is more due to the different disease manifestations they study and treat rather than imaging. We perform the biomarker detection experiments as detailed in Section~\ref{subsec:Contrastive} using PRIME and TREX trials separately. Specifically, we showcase the performance of intra-trial vs inter-trial experiments. Intra-trial refers to within PRIME and within TREX experiments - train and test within respective trials. Inter-trial refers to training and testing on different trials. The results are shown in Table~\ref{tab:Trial_domain}. The best results are obtained when training and testing on TREX. This is because of a larger diversity in TREX training data due to larger clinical trial window of 3 years. The eyes in TREX have more severe conditions than those present in PRIME. For this reason, combining the two dataset allows for a more complete distribution in terms of the severity of the disease and creates a more complete study and better results as shown in Table~\ref{tab:main_table}. Interestingly, the inter-trial results when training on TREX and testing on PRIME is higher than intra-trial training and testing on PRIME validating the need for larger diversity.

\paragraph{Domain shift before and after treatment}
The temporal element of the treatments studied in this dataset organically creates a domain shift between first and last visit's data for the same patient. We test this in Table~\ref{tab:Treatment_domain}. The experimental setup is biomarker detection from Section~\ref{subsec:Contrastive}. From the results in Table~\ref{tab:Treatment_domain}, it is clear that there is a domain difference between the first and last visits based on the inter vs intra-visit training and testing setups. However, the results of the last visit maybe skewed because of the effects of treatment that may cause improvement in some and deterioration/no change in others that might cause intra-visit variation in the last visit thereby leading to lower results. Hence, we further conduct domain adaptation experiments for this modality in Appendix~\ref{app: domain adaptation} in the coarser setting of DR/DME detection.

\begin{figure}[h!]
\centering
\includegraphics[scale = 0.6]{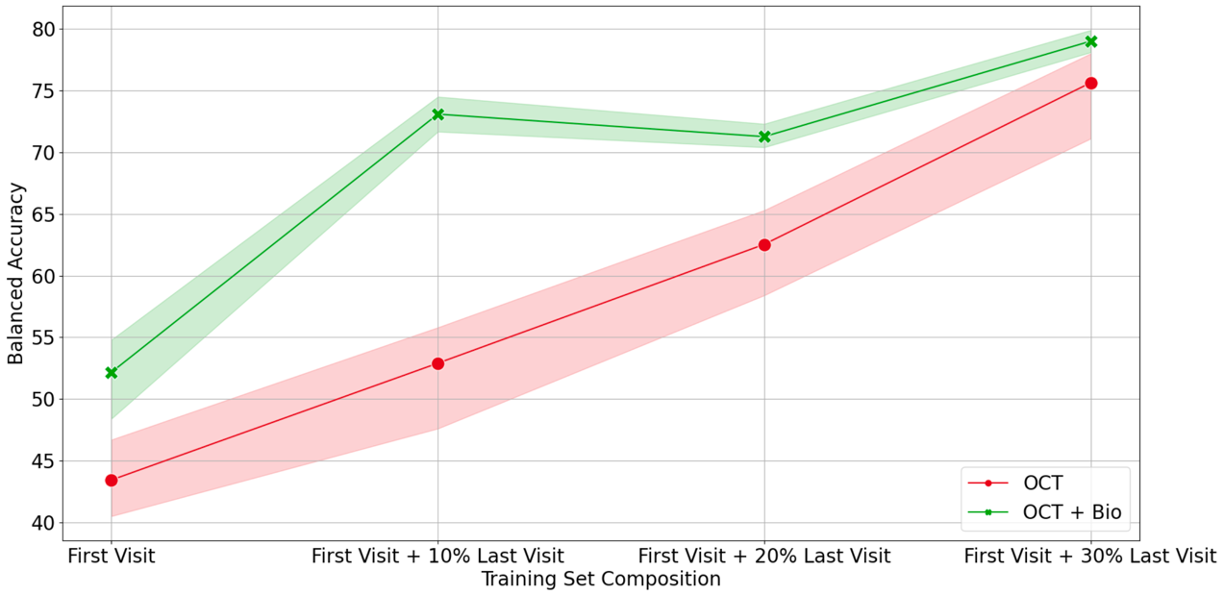}
\caption{A comparison between domain adaptation experiments using uni-modal and multi-modal data.}\vspace{-1.5mm}\label{fig:domainadapt}
\end{figure}

\subsection{Domain Adaptation before and after treatment}
\label{app: domain adaptation}
In this section, we study the domain adaptation between the first and last week's worth of data in terms of DR/DME detection. With that in mind, we train models on four different training sets and use a fixed test set to evaluate the transfer of knowledge between domains. The first training set consists of OCT collected solely from the patients initial visit to the clinic. The remaining training sets consist of OCT from the first visit and 10/20/30 \% of the OCT from the last visit respectively. The test set consists of the remaining 70\% of OCT from the final visit. There is no overlap between train and test sets. Training is repeated three times with different seeds and an average balanced accuracy and standard deviation is noted. To compare the effect of multiple modalities on domain adaptation, the same experiments are repeated using OCT and biomarkers.

In Fig \ref{fig:domainadapt}, the \texttt{x-axis} represents the composition of the training set used and \texttt{y-axis} the balanced accuracy achieved on the test set. Shaded regions around each curve show standard errors. The red curve highlights performance of OCT while the green curve shows the performance of OCT with biomarkers. We see that using data from the first visit only, results in the lowest performance for both curves. This training set has the largest disparity in its feature space compared to the test set. Conversely, the training set that combines 30\% of data from the final visit with the first visit achieved the highest balanced accuracy for both curves due to higher similarity existing between the source and target domains.

\subsection{Incorporation of Other Datasets alongside OLIVES}
\input{Tables/severity_score_results}

One of the useful features of the OLIVES dataset is that other medical datasets can be used in conjunction to develop other novel tasks. We incorporate the large amount of readily available healthy images from the Kermany dataset \cite{kermany2018labeled}  to train an auto-encoder which is later utilized to generate anomaly scores on the unlabeled data in the OLIVES dataset. Results from using this setup is shown in Table \ref{tab:severe_table} and the proposed strategy is identified with Kermany + OLIVES. We then use this anomaly score similar to a clinical label, within the contrastive learning setup in Section~\ref{subsec:Contrastive}. We observe that leveraging this information out-performs standard state of the art contrastive learning strategies, but doesn't out-perform the clinical contrastive learning with respect to multi-label AUROC we show in Table~\ref{tab:main_table}. This example demonstrates the adaptability of the OLIVES dataset through its potential to leverage information in other datasets to develop novel perspectives that didn't exist in the OLIVES dataset originally.

\subsection{Computational Resources}
\label{app: cluster}
All experiments were run on PCs with two NVIDIA GeForce GTX TITAN X 12 GB GPUs.

%% file: Tables/ContrastError.tex
\begin{table*}[h!]
\centering
\scriptsize
\begin{tabular}{ccccccc}
\toprule
\multirow{2}{*}{Method} & \multicolumn{6}{c}{Biomarkers}  \\ \cmidrule(l){2-7} 
% \\ \midrule
 & \multicolumn{1}{c}{IRF} & \multicolumn{1}{c}{DME} & \multicolumn{1}{c}{IRHRF} & \multicolumn{1}{c}{FAVF} & \multicolumn{1}{c}{PAVF} & \multicolumn{1}{c}{AUROC} \\
 \hline \midrule
PCL  \cite{li2020prototypical}              & \textbf{76.50}\% $\pm$ .513   & 80.11\% $\pm$ .335  & 59.1\% $\pm$ 1.03 & 76.30\% $\pm$ .378 & 51.40\% $\pm$ .556  & .767 $\pm$ .0017  \\

SimCLR \cite{chen2020simple}                    & 75.13\% $\pm$ .529  & 80.61\% $\pm$ .837  & 59.03\% $\pm$ 2.54  & 75.43\% $\pm$ .378  & 52.69\% $\pm$ 2.68 & .754 $\pm$ .0017\\
Moco v2 \cite{chen2020improved}                   & 76.00\%  $\pm$ .305     &  82.24\%  $\pm$ 1.38      & 59.6\%  $\pm$ .702      &  75.00\%   $\pm$ .608    &  52.69\% $\pm$ .472   & .770 $\pm$ .0035 \\ \hline \midrule

Eye ID                     & 72.63\% $\pm$ .264  & 80.2\%  $\pm$ .384  & 58\% $\pm$ 2.56   & 74.93\%  $\pm$ 1.36  & \textbf{65.56}\% $\pm$ .200  & .767 $\pm$ .0005 \\
CST                        & 75.53\%  $\pm$ .608  & \textbf{83.06}\% $\pm$ .213    & \textbf{64.3}\% $\pm$ 2.57  & 76.13\% $\pm$ .264   & 62.16\% $\pm$ 1.47  & .790 $\pm$ .0006\\
BCVA                       & 74.03\% $\pm$ .351  & 80.27\% $\pm$ .853  & 58.8\% $\pm$ 1.82  & \textbf{77.63}\% $\pm$ .305  & 58.06\% $\pm$ 1.27  & .776 $\pm$ .0017  \\

 \bottomrule
\end{tabular}
\caption{We show the performance of supervised contrastive training on the OLIVES dataset. In this table we explicitly show the standard deviation for the average across three runs for both accuracy and AUROC.}
\label{tab:main_std}
\end{table*}

%% file: Tables/Domain_Shift_Trial.tex
\begin{table}
  %\tiny
  \centering
  \caption{Benchmark results for characterizing domain shifts for data arising from the PRIME or TREX DME clinical trials.}
  %\caption{1}
  \label{tab:Trial_domain}
  \begin{tabular}{lcccc}
    \toprule
    Training Set & Test Set & Multi-Label AUROC \\ 
    \midrule
    Prime & TREX & .649 $\pm$ .024 \\
      TREX & Prime & .547 $\pm$ .013 \\
      Prime & Prime & .599 $\pm$ .042 \\
      TREX & TREX & .727 $\pm$ .011 \\
      \bottomrule
  \end{tabular}
\end{table}

%% file: Tables/Domain_Shift_Treatment.tex
\begin{table}
  %\tiny
  \centering
  \caption{Benchmark results for characterizing domain shifts before and after treatments in terms of first and last patient visits.}
  \label{tab:Treatment_domain}
  %\centering
  \begin{tabular}{lccccr}
    \toprule
    Training Set & Test Set & Multi-Label AUROC \\ 
    \midrule
First Visit & Last Visit & .628 $\pm$ .023 \\
      Last Visit & First Visit & .678 $\pm$ .012\\
      First Visit & First Visit & .712 $\pm$ .026  \\
      Last Visit & Last Visit & .546 $\pm$ .018\\
    \bottomrule
  \end{tabular}
  %\quad 
\end{table}

%% file: Tables/severity_score_results.tex
\begin{table}[h]
\centering

\centering
  \caption{Performance of leveraging data from a healthy dataset for a novel contrastive learning task is indicated by the Kermany + OLIVES row. Multi-Label is the average AUC from the multi-label classification task.}
\tiny
\begin{tabular}{@{}ccccccc@{}}

\toprule
\multicolumn{7}{c}{Severity Label Training Results (Accuracy / F1-Score)}   

\\ 
\midrule
\midrule
\multicolumn{1}{c}{Method} & \multicolumn{1}{c}{IRF} & \multicolumn{1}{c}{DME} & \multicolumn{1}{c}{IRHRF} & \multicolumn{1}{c}{FAVF} & \multicolumn{1}{c}{PAVF} & \multicolumn{1}{c}{Multi-Label} \\ \midrule
SimCLR \cite{chen2020simple}                    & 75.13\% / .715  & 80.61\% / .772 & 59.03\% /.675 & 75.43\% / .761 & 52.69\% / .249 & .754  \\
PCL \cite{li2020prototypical}                    & \textbf{76.50}\% / .717  & 80.11\% / .761 & 59.1\% /.683 & \textbf{76.30}\% / .773 & 51.40\% / .165 & .767  \\
Moco v2 \cite{chen2020improved}                   & 76.00\% / .720      &  82.24\% / .793       & 59.6\% / .692       &  75.00\% / \textbf{.784}      &  52.5\% / .201   & .769  \\ \bottomrule
%$SL_{100}$  & 75.16\% / .718 & 81.19\% / .783 & 61.13\% / .674 & 74.46\% / .752 & \textbf{56.86}\% / \textbf{.385} & .763  \\ 
%$SL_{1000}$ & 74.80\% / .714 & 81.73\% / .788 & 62.70\% / .694 & 72.26\% / .736 & 53.50\% / .280 & .763\\ 
Kermany + OLIVES & 75.20\% / .698  &81.46\% / .786 & \textbf{66.83}\% / .695 & 75.39\% / .756 & 54.7\% / .314 & \textbf{.774} \\

%\bottomrule
%$GC_{1000 + 10000}$ & 74.87\% / .716 & 80.82\% / .780 & 59.76\% / .686 & 74.73\% / .778 & \textbf{59.5}\% / \textbf{.431} & .771 \\
%$GC_{5000 + 10000}$ & 74.86\% / .722 & 82.48\% / .803 & 60.66\% / .697 & 71.56\% / .718 & 55.00\% / .315 & .759 \\
%$GC_{1000 + 5000 + 10000}$ & 73.57\% / .702 & 79.76\% / .762 & 61.13\% / .696 & 74\% / .746 & 57.53\% / .410 & .773 \\
 \bottomrule
\end{tabular}
\centering
\label{tab:severe_table}
\end{table}

%% file: Sections/appendixD.tex
\subsection{PRIME and TREX DME Clinical trials}\label{subsec:Trials}
\vspace{-1.5mm}
The PRIME \cite{hannah2021real} and TREX-DME \cite{payne2017randomized,payne2019randomized,payne2021long,wykoff2019intravitreal} clinical trials included at least 96 eyes with either center-involving diabetic macular edema (CI-DME, n = 56) or diabetic retinopathy without CI-DME (DR, n = 40) between December 2013 and April 2021. Each participant signed an informed consent form to participate in the clinical trial. Both trials were prospective, randomized clinical trials. Prospective trials refer to longitudinal studies that evaluate the outcome of a particular disease during treatment. In PRIME, 40 eyes with nonproliferative diabetic retinopathy (NPDR) or proliferative diabetic retinopathy (PDR) without CI-DME received intravitreal aflibercept injections (IAI) monthly until the eyes achieved a diabetic retinopathy severity scale (DRSS) score improvement of $\geq$ 2 steps; at baseline, eyes were randomized 1:1 into two management strategies for DR: 1) DRSS-guided or 2) panretinal leakage index (PLI)-guided management. In TREX-DME, 150 eyes with CI-DME were randomized 1:2:2 into three cohorts for management with ranibizumab (0.3 mg): 1) monthly treatment or 2) treat and extend, or 3) treat and extend with angiography-guided macular laser photocoagulation. For each patient, general demographics, ocular disease state data (e.g., best corrected visual acuity (BCVA)), central subfield thickness measurements (CST), and detailed ocular imaging (e.g., spectral-domain optical coherence tomography (SD OCT), fundus photography, and fluorescein angiography) was obtained per the protocol in Section~\ref{app: all_labels}. All SD-OCT images were obtained using the Heidelberg Spectralis HRA+OCT (Heidelberg Engineering, Heidelberg, Germany) with a volume-per-cube acquisition protocol (20 x 20, 49 lines, 768 A-scans per line) with 9-times image averaging.

\subsection{Clinical Study Process Description}
\label{app: trail_description}
The Intravitreal Aflibercept as Indicated by Real-Time Objective Imaging to Achieve Diabetic Retinopathy Improvement (PRIME) study was a prospective, randomized, phase II clinical trial (ClinicalTrials.gov identifier, NCT03531294; IND138997). The purpose of the study was to assess the safety and efficacy of as-needed intravitreal aflibercept injections for eyes with diabetic retinopathy without center-involved diabetic macular edema via the guidance of real-time Diabetic Retinopathy Severity Scale (DRSS) level or panretinal leakage index (PLI) assessment. The DRSS level was determined by color fundus photography graded by a trained image analyst. PLI assessment was conducted by an automated ultrawidefield fluorescein angiography image analysis platform. Between May 2018 and March 2019, forty subjects were enrolled in PRIME given the following inclusion criteria: 18 years of age and older with type 1 or type 2 diabetes mellitus, a DRSS level of 47A to 71A as determined by the CRC (Cole Eye Institute, Cleveland Clinic, Cleveland, Ohio, USA), and Early Treatment Diabetic Retinopathy Study (ETDRS) best-corrected visual acuity (BCVA) of 20/800 or better. The exclusion criteria consisted of CST greater than 320 $\mu$m in the study eye; central DME causing vision loss; vitreous hemorrhage; previous treatment of anti-vascular endothelial growth factor (VEGF) pharmacotherapies, corticosterids, dexamethasone, or fluocinolone acetonide in the study eye; and a history of vitrectomy or panretinal photocoagulation. Further details are available in~\cite{hannah2021real}.

The Treat and Extend Protocol in Patients with Diabetic Macular Edema (TREX-DME) study was a prospective, randomized, phase I/II, multicenter clinical trial (ClinicalTrials.gov identifier, NCT01934556). The purpose of the study was to compare the administration of intravitreal ranibizumab injections for eyes with center-involving diabetic macular edema on the basis of monthly dosing or a treat and extend algorithm with and without angiography-guided macular laser photocoagulation. Between November 2013 and April 2015, 150 eyes from 116 subjects were enrolled in TREX-DME given the following inclusion crtieria: type 1 or type 2 diabetes mellitus, center-involving DME, ETDRS BCVA between 79 and 24 letters (Snellen equivalent, 20/25-20/320). The exclusion criteria consisted of previous treatment of anti-VEGF pharmacotherapies, corticosteroids, or focal macular laser. Fifty-six out of the 150 study eyes were evalulated at Retina Consultants of Texas study sites and are included in this data set. Further details are available in~\cite{payne2017randomized,payne2019randomized,payne2021long,wykoff2019intravitreal}.

A summary of the two studies and the processes involved is provided as Summary-DR-DME-Studies.docx under the labels folder accessed through Appendix~\ref{app: Access}.

\begin{figure}[t]
\centering
\includegraphics[scale = .38]{Images/PRIME.pdf}
\vspace{-1.5mm}
\caption{ML Centric Labels Datasheet within the OLIVES Dataset for PRIME trial}\vspace{-1.5mm}\label{fig:Labels_Screen_PRIME}
\end{figure}

\begin{figure}[t]
\centering
\includegraphics[scale = .38]{Images/TREX.pdf}
\vspace{-1.5mm}
\caption{ML Centric Labels Datasheet within the OLIVES Dataset for TREX-DME trial}\vspace{-1.5mm}\label{fig:Labels_Screen_TREX}
\end{figure}

\subsection{ML Centric Label description}
\label{app: ML_datasheet}
The ml centric labels directory within the label access provided in Appendix~\ref{app: Access} consists of two CSV files. The first is the Biomarker-Clinical-Data-Images.csv. The labels for all $9408$ images with biomarker alebls are provided in the csv file. Two screenshots of this CSV file from both the PRIME and TREX-DME trials are shown in Fig.~\ref{fig:Labels_Screen_PRIME} and~\ref{fig:Labels_Screen_TREX} respectively. The first column provides a path within the image folder structure. The path includes the following:
\begin{enumerate}
    \item Trial: Can refer to either PRIME or TREX-DME.
    \item Arm: TREX-DME has an additional cohort-based subfolder within the trial: GILA, Monthly, and TREX that identify specific cohorts of patients based on treatment.
    \item Folder: Refers to the code that identifies each patient.
    \item Visit: Refers to the visit that the current images and labels refer to. Note that in both the clinical studies, the biomarkers are retrospectively added to the first and last visits. Hence, in TREX-DME, the biomarkers are labeled at V1 and V22 for the first considered patient and so on. 
    \item Eye: The possible values are "OD" or "OS" and this serves to identify the right or left eye, respectively. 
    \item Image name: The name that is provided in the dataset directory on Zenodo.
\end{enumerate}
Scan can be one of $49$ slices that exists in a 3D volume which is obtained from the OCT machine for every patient for every visit. The next $16$ columns refer to biomarkers the full list of which is present in Table~\ref{tab:olives} and whose generation process and abbreviations are described in Appendix~\ref{app: biomarkers}. $1$ indicates their presence for the considered scan while $0$ indicates their absence.

The last four columns refer to clinical labels - Eye ID, BCVA, CST, and Patient ID. The ranges of BCVA and CST are shown in Fig.~\ref{fig: distribution} while their significance is expanded in Appendix~\ref{app: labels}.

The second csv file under ml centric labels is Clinical-Data-Images.csv. This file holds the path name and only the four clinical labels for all $78,189$ scans.